\documentclass[review]{elsarticle}

\usepackage{hyperref}
\usepackage{amssymb, graphics, setspace}
\usepackage{pgfplots}
\usepackage{textcomp}
\usepackage[caption=false]{subfig}
\usepackage{amsmath}
\usepackage{commath}
\usepackage{graphicx,bm}
\usepackage{url}
\usepackage{float}
\usepackage{textcomp}
\usepackage{verbatim}
\usepackage{multirow}

\usepackage{url}
\usepackage{ulem}

\begin{document}
\begin{frontmatter}

\title{Pomeron/glueball and odderon/oddball trajectories}


\author[a]{Istv\'an Szanyi}
\ead{istvan.szanyi@cern.ch}
\address[a]{E\"otv\"os Lor\'and University,\\ H-1117 Budapest, P\'azm\'any P\'eter s., 1/A HUNGARY}

\author[b]{L\'aszl\'o Jenkovszky}
\ead{jenk@bitp.kiev.ua}
\address[b]{Bogolyubov Institute for Theoretical Physics (BITP),\\
	Ukrainian National Academy of Sciences \\14-b, Metrologicheskaya str.,
	Kiev, 03680, UKRAINE}

\author[c]{Rainer Schicker}
\ead{schicker@physi.uni-heidelberg.de}
\address[c]{Physikalisches Institut, University Heidelberg, GERMANY}

\author[d]{Volodymyr Svintozelskyi} 
\ead{1vladimirsw@gmail.com}
\address[d]{T.H. Shevchenko Kyiv Nat. University, UKRAINE}

\begin{abstract}
	  We predict glueball/oddball resonances lying on the pomeron/odderon trajectories.
	A simple new form of the trajectories, with threshold and asymptotic
	behaviour required by analyticity and unitarity, is proposed.
	The parameters of these (pomeron and odderon) trajectories are fitted to the data on
	high-energy elastic proton-proton and proton-antiproton scattering. The fitted trajectories are
	extrapolated to the resonance region to predict masses and widths of
	glueballs and oddballs. The (pomeron and odderon) trajectories may be used to
	calculate processes of central exclusive diffraction (CED).   
\end{abstract}

\end{frontmatter}


\section{Introduction} \label{s1}

Regge trajectories $\alpha(t)$ connect the scattering region, $t<0$, with that of particle spectroscopy, $t>0$. Discussing Regge trajectories we use a single variable $t$ for both its positive and negative range. In this way we realize crossing symmetry and anticipate duality: dynamics of two kinematically disconnected regions is interrelated: the trajectory at $t<0$ "knows" its behaviour in the cross channel and vice versa. Most of the familiar meson and baryon trajectories follow the above regularity: with their parameters fitted in the scattering region they fit masses and spins of relevant resonances (see, \textit{e.g.} Refs.~\cite{Collins,  BP, DDLN}). The behaviour of trajectories both in the scattering and particle region is close to linear, with parameters (intercepts and slopes) compatible with fits in the scattering region.
 This observation, combined with the properties of dual models and hadron strings resulted in a prejudice of the linearity of Regge trajectories, although it is obvious that resonances' widths on the one hand, and unitarity on the other hand, are incompatible with real linear Regge trajectories. 
 
 In this paper, we concentrate on the study of the pomeron trajectory with
 even C-parity, and its C-odd counterpart, the odderon trajectory.
 The nature of the odderon, and the behaviour of its trajectory, are
 still shrouded in mystery \cite{Ewerz:2003xi,Ewerz:2005rg,Ewerz:2013kda}.
 The experimental signatures of odderon exchanges continue to be hotly
 debated within the community \cite{Martynov:2017zjz,Khoze:2018bus,Szczurek:2019kym,Csorgo:2018uyp}.
 
From the simple, "canonical"  pomeron trajectory \cite{DDLN} $\alpha(t)=1.08+0.25t$, the mass of the lightest glueball with $J=2^{++}$ is $M=\sqrt t=1.92$ GeV. Experimental evidence of this glueball state is so far missing. There are many alternative predictions,  with more sophisticated derivation of the pomeron and odderon trajectories from advanced theories such as supergravity \cite{Caceres,Caceres2}, AdS supergravity duality \cite{Brower:2000rp}, QCD (lattice) \cite{2,LlanesEstrada:2005jf,Kaidalov,Meyer,4}, and gauge/string duality \cite{Boschi-Filho, Ballon-Bayona:2015wra, Brunner:2015oqa, Capossoli:2016ydo, Rodrigues}. 

We do not specify the gluonic content of the glueball/oddball states, pointing out that they are presumed resonances made of an even/odd number of gluons, lying on the pomeron/odderon trajectory. These states may be more complicated hybrid states, containing also quarks and antiquarks.

 Fits to scattering data leave little room for variations of the parameters (and predictions of glueball masses). Also note that the real and linear trajectories do not contain information on decay widths of the glueballs indicating the importance of nonlinear complex trajectories. 

The crucial role of the imaginary part of the trajectory and its relation to analyticity were always implied, but the problem is still open. Disputed is even the sign of the curvature (convex or concave?) in the trajectories (see, {\it e.g.} Ref.~\cite{Chen}). Among first steps toward an explicit solution were made in Ref.~\cite{Predazzi}.
Analyticity and unitarity impose \cite{Gribov,Barut,DAMA} severe constraints on the threshold behaviour of the
trajectories. Combined with the asymptotic bounds, they provide reasonable  constraints to proceed with model building.

The approximate linearity of Regge trajectories is an indication of non-perturbative, string-like dynamics in low-$|t|$ scattering of hadrons, difficult to treat in the framework of quantum chromodynamics (QCD) \cite{Kaidalov:2005kz}. It was realized in the late 1960's  that the properties of the hadronic interactions are well described by string-string interactions, where hadrons are identified as vibrational modes of the quantum strings. Strings do not have point-like interaction, thus they do not result in ultraviolet divergences.
	
A non-perturbative method, applicable in QCD for description of large distance dynamics, is the $1/N$-expansion (called also topological expansion). In this approach the quantities $1/N_c$ or $1/N_f$, where $N_c$ is the number of colours and $N_f$ is the number of light flavours, are considered as small parameters, while amplitudes and Green functions are expanded in terms of these quantities. In the formal limit $N_c\rightarrow\infty,\ \ N_f/N_c\rightarrow 0$. One hopes to obtain an exact solution of the theory in this limit (2-dimensional QCD has been solved in the limit $N_c\rightarrow\infty$. However this approximation is rather far from reality, as resonances in this limit are infinitely narrow. More realistic is the case when $N_f/N_c\sim 1$ is fixed and the expansion is in $1/N_f$ (or $1/N_c$), called topological expansion \cite{Veneziano:1974fa, Veneziano:1976wm}, may be more realistic. Note that the $1/N$ expansion is applicable to Green-functions or amplitudes for white states.
	
Despite tremendous efforts and impressive successes, in field-theoretical approaches, namely QCD and string theories, the Regge-pole theory remains the main practical tool to handle soft scattering processes \cite{Kaidalov:2005kz}. Its  efficiency has motivated efforts to improve the model in two related directions: 1) derive Regge-pole behavior from “first principles” \textit{i.e.} from QCD and 2) to reduce the remaining freedom in the model, in particular the form of the Regge trajectories and their parameters. The leading role in this development belongs to late L. Lipatov's group, authors of a large number of widely cited papers \cite{Fadin:1975cb,Lipatov:1976zz,Kuraev:1976ge,Kuraev:1977fs,Balitsky:1978ic}, in which they try to derive from perturbative QCD the form of the singularity, that turns out to be a infinite number of poles in the angular momentum plane, accumulating towards its rightmost point. It may be approximated with a cut. The intercept of the vacuum (pomeron) trajectory is “supercritical”, \textit{i.e.} its value is greater than one (approximately 1.3 but its exact value depends on the approximations made).  Asymptotically the pomeron trajectory is logarithmic. Remind that an important premise in the derivation of the BFKL pomeron is gluon reggeisaton, \textit{i.e.}, the Regge behaviour (of gluons, constituens of the pomeron) is built in from the beginning.  Due to its perturbative nature, the credibility of the BFKL pomeron increases with $s$ and $|t|$.
	
Much more ambitious are later developments connected with string models. Modern string theory grew out of the failed attempt to treat strong interactions as a string theory. While the approach had various phenomenological successes, it was ultimately abandoned as the fundamental theory of strong interactions.
	
With the emergence of QCD as a viable field theory for strong interactions,  string theory reemerged later as a “theory of everything”. Phenomenological successes of the hadronic string picture: string theories naturally give rise to linear Regge trajectories (Chew-Frautchi plot) as well to a Hagedorn spectrum for the density of hadrons $\rho_m\sim \exp(m/m_H)$, where $m_H$, the Hagedorn temperature, is a mass parameter. Thermodynamically it corresponds to an upper bound on the temperature of hadronic matter. Strings are assumed to emerge in QCD as flux tubes.
	
There is a compatibility problem in unifying or comparing two different approaches to forward physics, namely that based on the analytical $S$-matrix theory, namely its realization within the Regge-pole model, and the underlying standard model theory within QCD and/or the string model. When switching between the scattering $t<0$ and spectroscopy $t>0$ kinematical domains, one necessarily passes through the $t=0$ point, a very special point from strong-coupling dynamics. It is a well-known problem that QCD coupling constant, for example, is undetermined in this region. From the viewpoint of confinement, a small vicinity of $t=0$ (infrared limit) is problematic primarily due to yet unknown transition between the dynamics of QCD partonic (coloured) and hadronic (colorless) degrees of freedom. Such regime associated with hadronisation is not predicted from the first principles, but rather successfully modelled in the framework of string model. At macroscopic distances corresponding to this special limit $t\rightarrow0$, long QCD strings break up creating shorter strings resembling hadrons. Apparently, string break ups have nothing to do with analyticity at $t=0$. 
Confinement and analyticity are not compatible notions in QCD in the strongly coupled IR regime of small $|t|$: they are rather contradictory to each other; and yet the old-fashioned analyticity arguments are being persistently used for constraining the properties of amplitudes and Regge trajectories. When continuing the scattering regime $t<0$ into the resonance domain $t>0$, one has to deal with confinement and associated breakdown of analyticity.

This paper is organized as follows. In Sec.~\ref{sec:unandu}, constraints on the trajectory based on unitarity, analyticity and duality are discussed. The use of pomeron/odderon trajectories in central exclusive production is outlined in Sec.~\ref{sec:POced}.  A simple analytic Regge trajectory is presented in Sec.~\ref{sec:trajectory}. Single and double pole Regge fits to elastic scattering data are described in Sec.~\ref{sec:SP} and Sec.~\ref{Sec:DP}. Pomeron-pomeron and odderon-odderon total cross sections are derived in Sec.~\ref{Amplitude}. Masses and widths of predicted glueball and oddball states are presented in Secs.~\ref{sec:glueballs} and \ref{sec:oddballs}, and conclusions are drawn in Sec.~\ref{sec:conclusions}.

\section{Unitarity, analyticity and duality constraints on trajectories}\label{sec:unandu}

Unitarity imposes \cite{Barut} a severe constraint on the threshold behaviour of the trajectories:  
\begin{equation} \label{thr}
\Im m\alpha(t)_{t \rightarrow t_0}\sim(t-t_0)^{\Re e\alpha(t_0)+1/2},
\end{equation}
while asymptotically the trajectories are constrained by \cite{DAMA}
\begin{equation}\label{asympt}
\Bigg\vert\frac{\alpha(t)}{\sqrt{t}\ln{t}}\Bigg\vert_{t\rightarrow\infty}\leq {\rm const}.
\end{equation}
The above asymptotic constraint can be still lowered to a logarithm by imposing (see Ref.~\cite{Rivista} and earlier references) wide-angle power behaviour for the amplitude. 
    
    The above constraints are restrictive but still leave much room for model building. In Refs. \cite{Francesco_m, Francesco_b} the imaginary part of the trajectories (resonances' widths) was recovered form the nearly linear real part of the trajectory by means of dispersion relations and fits to the data. 
    
Models of trajectories satisfying constraints by unitarity, analyticity and duality \cite{DAMA} can be found in Refs. \cite{Burak,Burak2,2}.
        
While the parameters of meson and baryon trajectories can be determined both from the scattering data and from the particles spectra, this is not true for the pomeron (and odderon) trajectory, known only from fits to scattering data (negative values of its argument). An obvious task is to extrapolate the pomeron trajectory from negative to positive $t$-values to predict glueball states at $J=2, 4,...$, for which, however, no experimental evidence exists so far. Given the nearly linear form of the pomeron trajectory, known from the fits to the (exponential) diffraction cone, little room is left for variations in the region of particles ($t>0$). 

The non-observation so far of any glueball state in the expected values of spin and mass may have two explanations:

   \begin{itemize}
   \item glueballs appear as hybrid states mixed with quarks that makes their identification difficult;
     \item the production cross section $d\sigma/dt$ for glueballs is low, and their width is large.
       \end{itemize}
   
To resolve these problems one needs a reliable model to predict cross sections and decay widths of the expected glueballs in which the pomeron trajectory plays a crucial role.
                   


The pomeron exchange dominates the diffraction cone, whose shape deviates from exponential in three cases: 

\begin{enumerate}
\item A smooth curvature up to the exponential cone around $t= -0.1$ GeV$^2$, where the slope of
the cone changes by about $2$ units of GeV$^{-2}.$ This phenomenon (the so called "break"), physically related to nucleon's
"atmosphere" and required by $t$-channel unitarity, comes from the threshold singularity in the amplitude,
encoded (mainly) in the Regge trajectory and (partly) in the Regge residue (see Ref.~\cite{Tan} and earlier references
therein).

\item The dip, slowly (logarithmically) moving from $t\approx -1.3$ GeV$^2$
at $\sqrt{s} \sim$ 50 GeV (ISR) to $t\approx -0.5$ GeV$^2$ at $\sqrt{s} \sim$ $13$ TeV,
is a property of the amplitude rather than the trajectory, nearly linear in this region. In other words, the dramatic dip-bump structure (diffraction minimum and maximum) does not affect the nearly linear behaviour of the trajectories in that, around $-1$ GeV$^2$, interval.    

\item The apparent slow-down of the exponential decrease of the cross section beyond  $t\approx -10$ GeV$^2$ is due to transition from "soft" to "hard" dynamics and it can be mimicked \cite{Jenk1} by the gradual slow down (from linear to logarithmic) rise of the pomeron and odderon trajectories. 

\end{enumerate}  
  
Whatever the details of the parametrization, the relative role of the odderon increases with $|t|$ just because its slope is smaller than that of the pomeron trajectory. The odderon is barely visible near $t=0$, causing endless disputes, but its role is crucial at the dip and remains important beyond it. This motivates our inclusion of the odderon in our fits to elastic scattering data, on the one hand, but, on the other hand, it enables us to make predictions for the three-gluon (oddball) states.  

\section{Pomeron/odderon trajectories in Central Exclusive Production}\label{sec:POced}

Central exclusive diffractive (CED) production continues to attract the attention of both theorists and experimentalists (see, \textit{e.g.} Refs.~\cite{Schicker2,Ewerz:2019arb} and references therein). This interest is triggered by LHC's high energies, where even the subenergies at equal partition are sufficient to neglect the contribution from secondary Regge trajectories and consequently CED can be considered as a gluon factory to produce exotic particles such as glueballs. 

\begin{figure}[H] 
	\centering
	\includegraphics[scale=0.5]{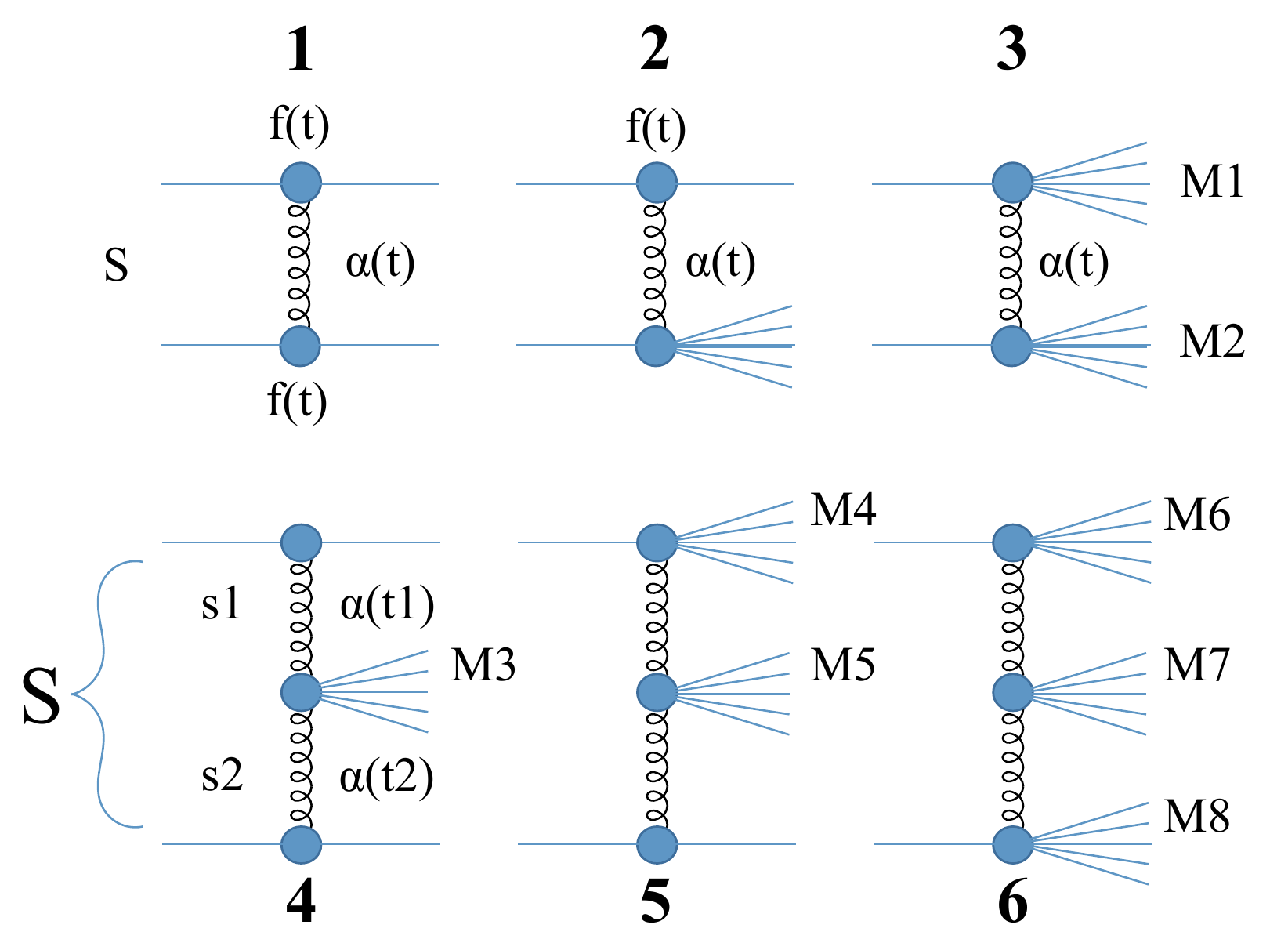}
	\caption{Regge-pole factorization.}
	\label{fig:1}
\end{figure}

The Regge-pole factorization is shown in  Fig.~\ref{fig:1}. In  single-diffraction dissociation or single dissociation (SD) one of the incoming protons dissociates, in double-diffraction dissociation or double dissociation (DD) both protons dissociate, and in central diffraction (CD) or double-Pomeron exchange (DPE) neither proton dissociates. These processes are tabulated below,
\begin{eqnarray}
{\rm SD}& pp\rightarrow p^*p\nonumber\\
{\rm or}&pp\rightarrow pp^*\nonumber\\
{\rm DD}&pp\rightarrow p^*p^*\nonumber\\
{\rm CD}\;{\rm (DPE)}&pp\rightarrow pXp,\nonumber
\label{eqn:processes}
\end{eqnarray}
where $p^*$ represents a diffractively dissociated proton and $X$ denotes a central system, consisting of meson/glueball resonances. 

Below we study CED represented by topology 4. Its knowledge will be essential in future studies with diffractively excited protons, represented by topologies 5 and 6.

 The basic sub-process is pomeron-pomeron scattering, producing glueballs lying on the direct-channel pomeron trajectory, with a triple pomeron vertex. Glueball “towers”, \textit{i.e.} excited glueball states, called reggeized Breit-Wigner resonances, lie on this trajectory. Crucial for the identification of these states is knowledge of the non-linear complex trajectory, interpolating between negative and positive values of its argument. While the real part of the trajectory is almost linear, the recovery of the imaginary part, determining the widths of predicted glueballs, is a highly non-trivial problem. 
 
We continue the lines of research initiated in Refs. \cite{Schicker1, Schicker2} in which an analytic pomeron trajectory was used to calculate the pomeron-pomeron cross section in central exclusive production measurable in proton-proton scattering. In the present work we introduce a new model of non-linear complex Regge trajectory, both for the pomeron and odderon, satisfying the requirements of the analytic $S$-matrix theory. We first fit the parameters of those trajectories to high-energy elastic proton-proton and proton-antiproton scattering data then extrapolate the fitted new trajectories to the particle region to predict the masses and widths of the glueballs and oddballs lying respectively on the pomeron and odderon trajectories.

\section{Simple analytic Regge trajectory}\label{sec:trajectory}

What is the simplest ansatz for a Regge trajectory satisfying the following constraints:

\begin{itemize}
\item threshold behaviour imposed by unitarity, Eq.~(\ref{thr}),
\item asymptotic behaviour constrained by Eq.~(\ref{asympt}),
\item yet compatible with the nearly linear behaviour in the resonance region (Chew-Frautschi plot)?
\end{itemize}

Attempts and explicit examples can be found in a number of papers (see, \textit{e.g.} Refs.~\cite{Schicker1, Schicker2} and earlier references therein).

The trajectory is:

\begin{equation}\label{3}
    \alpha(t) = \frac{1+\delta+\alpha_{1}t}{1+\alpha_{2}\big(\sqrt{t_{0}-t}-\sqrt{t_0})},
\end{equation}
where $t_0=4m_{\pi}^2$ for pomeron and $t_0=9m_\pi^2$ for odderon and $\delta, \alpha_1, \alpha_2$ are parameters, real and of
positive value, to be fitted to scattering ($t<0$) data with the obvious constraints: $\alpha(0)\approx 1.08$ and
$\alpha'(0)\approx 0.3$ (in case of the pomeron trajectory). Trajectory Eq.~(\ref{3}) has square-root asymptotic behaviour, in accordance with the requirements of the analytic $S$-matrix theory. The novelty of this model for the Regge trajectory, compared to earlier ones \cite{Schicker1}, that it produces physical resonances (glueballs) whose widths are widening as their masses increase.

With the parameters fitted in the scattering region, we continue trajectory Eq.~(\ref{3}) to positive values of $t$. When approaching the branch cut at $t=t_0$ one has to choose the right Riemann sheet. For $t>t_0$ trajectory Eq.~(\ref{3}) may be rewritten as     
\begin{equation}\label{3m}
    \alpha(t)=\frac{1+\delta+\alpha_1t}{1-\alpha_2(i\sqrt{t-t_0}+\sqrt{t_0})},
\end{equation}
with the sign "minus" in front of $\alpha_{2}$, according to the definition of the physical sheet. 

For $t>>t_0,\ \  |\alpha(t)|\rightarrow\frac{\alpha_1}{\alpha_2}\sqrt{|t|}$. For $t>t_0$ (on the upper edge of the cut), $\Im m\alpha>0.$

The intercept is $\alpha(0)=1+\delta$ and the slope at $t=0$ is 
\begin{equation}\label{Slope}
\alpha'(0)=\alpha_1+\alpha_2\frac{1+\delta}{2\sqrt{t_0}}.
\end{equation}    

To anticipate subsequent fits and discussions, note that the presence of the light threshold $t_0=4m_{\pi}^2$ (required by unitarity and the observed "break" in the data) results in an increased, compared with the "standard" value of about $0.25$ GeV$^{-2}$, slope. 

The crucial task, on which glueball (and oddball) predictions are based is the correct fit of the pomeron (and odderon) trajectory to the data. The most direct and reliable way to determine the free parameters of the pomeron (and odderon) trajectory is fitting the high-energy elastic proton-proton and proton-antiproton scattering measurements. The data in the TeV energy range (TEVATRON, LHC) are the best for this purpose, where the contribution of secondary reggeons is negligible \cite{Szanyi}.

The situation, however, is not that simple. The smooth, nearly exponential small-$|t|$  part of the cone before the dip ($|t|\lesssim0.3$) at the LHC is too short to fit the trajectory and provide its reliable extrapolation to large positive values, where glueballs are expected. Fits in this limited interval are under control,
and the small deviation from an exponential of the cone can be parametrized by a pomeron exchange within the simple Donnachie-Landshoff model \cite{DDLN}, see next section, where the resulting trajectory inherits the curvature, called "break" seen both at the ISR and LHC near $t= -0.1$ GeV$^2$.
    
\section{Single pole Regge fit to low-$|t|$ elastic scattering data}\label{sec:SP}

In this section we introduce a simple single pole (SP) Regge model and fit its parameters to the LHC TOTEM 13 TeV low-$|t|$ elastic proton-proton scattering data. 

High-energy elastic proton-proton scattering data, including ISR and LHC energies, were successfully fitted with non-linear pomeron trajectories in number of papers (see Ref.~\cite{Tan} and references therein). 
Since here we are interested in the parametrization of the pomeron trajectory, dominating the LHC energy region, we concentrate on the LHC data, where secondary trajectories can be completely ignored.  

\begin{figure}[H] 
	\centering
	\includegraphics[scale=0.44]{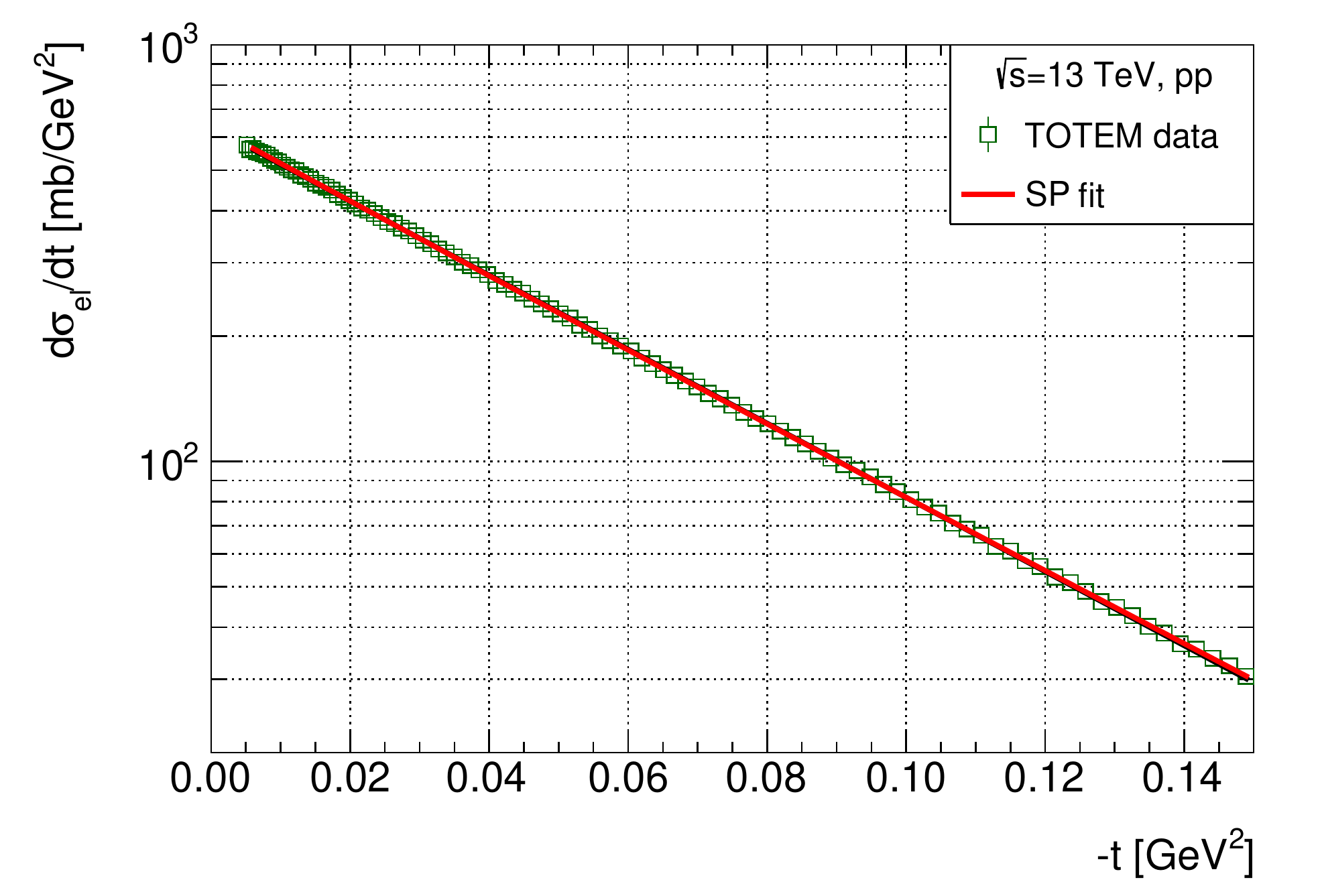}
	\caption{Fitted $pp$ TOTEM 13 TeV differential cross section data \cite{totem13} using the amplitude Eq.~(\ref{Eq:Pf}) and trajectory Eq.~(\ref{3}).}
	\label{fig:dsdtsimp}
\end{figure}

\begin{figure}[H] 
	\centering
	\includegraphics[scale=0.44]{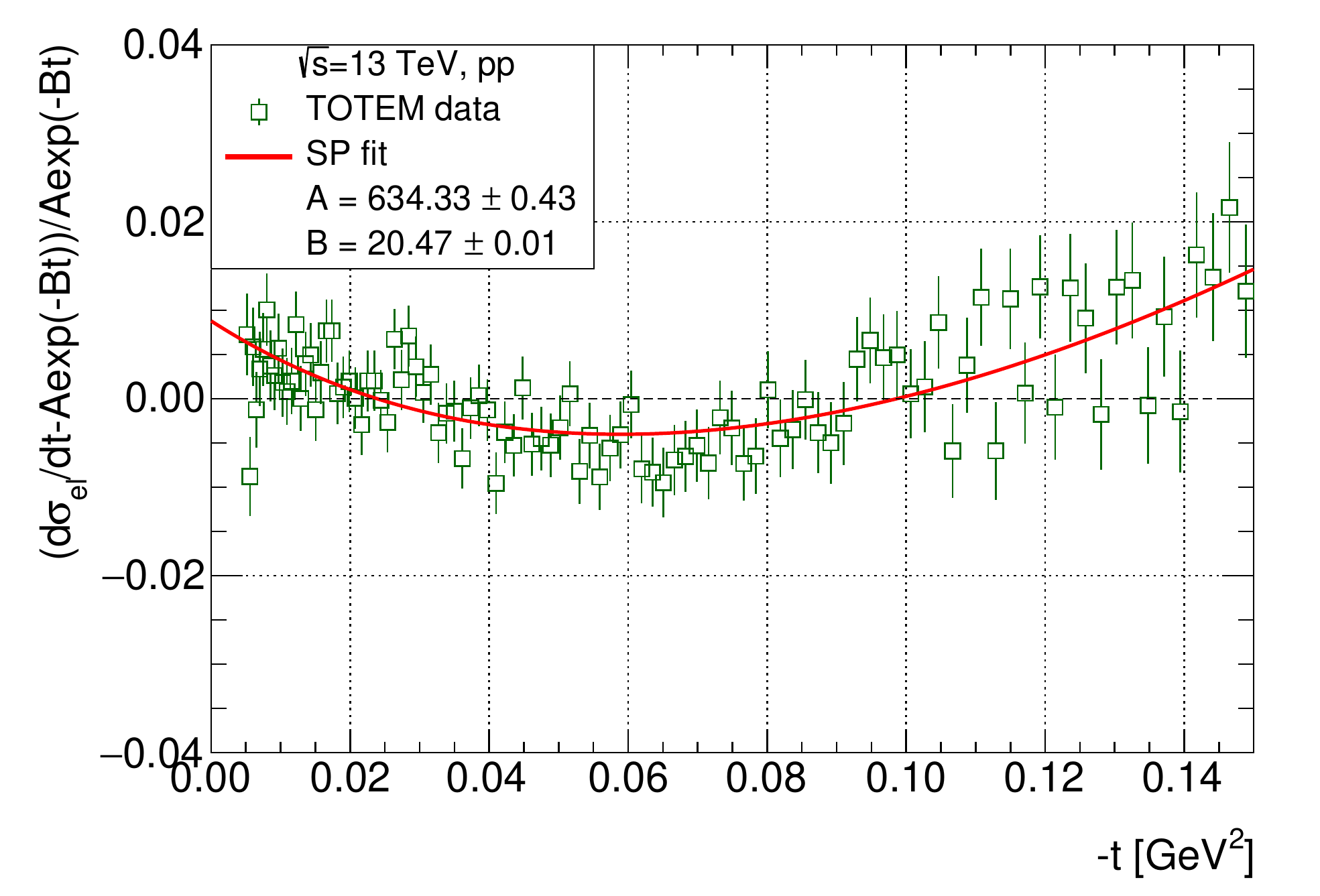}
	\caption{Fitted $pp$ TOTEM 13 TeV differential cross section data \cite{totem13} in normalized form using the amplitude Eq.~(\ref{Eq:Pf}) and trajectory Eq.~(\ref{3}).}
	\label{fig:dsdtsimp_n}
\end{figure}

Thus the amplitude in the SP model:       
\begin{equation}\label{Eq:Pf}
A_P(s,t)=a_Pe^{b_Pt}e^{-i\pi\alpha_P(t)/2}(s/s_{0P})^{\alpha_P(t)}.
\end{equation}
Using the norm,
\begin{equation}\label{Eq:Norm}
\frac{d\sigma_{el}}{dt}(s,t)=\frac{\pi}{s^2}|A_P(s,t)|^2,
\end{equation}
we fitted the TOTEM 13 TeV elastic $pp$ differential cross section data \cite{totem13} in the interval $0.005\leqslant|t|\leqslant0.15$ GeV$^2$, where for the pomeron trajectory, $\alpha_P(t)$, Eq.~(\ref{3}) was used. The values of the fitted parameters with fit statistics are shown in Tab.~\ref{tab:parametersDL_n}. The fitted differential cross section and its normalized form are shown in Figs.~\ref{fig:dsdtsimp} and \ref{fig:dsdtsimp_n}. The normalized form, used originally by TOTEM \cite{totem13}, shows the deviation of the low-$|t|$ diffraction cone from the purely exponential form.

\begin{table}[H]    
	\centering
	\caption{Values of the fitted parameters of the single pole Regge model for $pp$ 13 TeV data on elastic differential cross section. Values of the fitted parameters and their errors are rounded up to four valuable decimal digits.}
	\begin{tabular}{|cccc|cccc|}
		\hline\hline
		\multicolumn{4}{|c|}{ Pomeron }  & \multicolumn{4}{c|}{ Fit statistics }  \\
		\hline
		& $a_P~[\sqrt{\rm mbGeV^2}]$ & $5.018 \pm 0.0106$ & &&&& \\
		& $b_P$ & $3.931 \pm 0.040 $ & & & \multirow{2}{*}{ $\chi^2$ } & \multirow{2}{*}{ $100.77$ } & \\
		& $\delta_P$ & $0.08009 \pm 0.00011$ & & & \multirow{2}{*}{ $NDF$ } & \multirow{2}{*}{ $98$ } & \\
		& $\alpha_{1P}~[{\rm GeV}^{-2}]$ & $0.2980 \pm  0.0021$ & & & \multirow{2}{*}{ $\chi^2/NDF$ } & \multirow{2}{*}{ $1.03$ } & \\
		& $\alpha_{2P}~[{\rm GeV}^{-1}]$ & $0.02467 \pm 0.00128 $ & &&&& \\
		& $s_{0P}~[{\rm GeV}^2]$ & $1.0$ (fixed) & & & & & \\
		\hline\hline
	\end{tabular}
	\label{tab:parametersDL_n}
\end{table}

Note that, in this paper, we treat only the strong (nuclear) amplitude, separated from Coulombic forces. The CNI effects modify the nuclear cross-section by less than 1 \% for $|t|\gtrsim0.007$ GeV$^2$, thus,  in the nuclear range, the CNI effects can be ignored, making the fit independent of the interference modelling \cite{totem13}.      

To restore the trajectory in a wider interval (interesting from the point of view of glueballs!), the influence of the low-$|t|$ "break" must be compensated by fits to the larger $|t|$ region, that beyond the dip. We do it in Sec. \ref{Sec:DP} by using a dipole (DP) model that proved to be efficient in earlier studies \cite{Szanyi,JLL,PEPAN}. The SP model is applicable only at low-$|t|$, to reproduce the higher-$|t|$ features (dip-bump) of the experimental data a more developed model is required which includes also the odderon as its relative contribution increases with momentum transfer \cite{DDLN,Szanyi}. The presence of an odderon exchange allows us also to make predictions for the oddballs.

\section{Double pole Regge fit to elastic scattering data in the dip-region} \label{Sec:DP}

While at lower energies, \textit{e.g.} at the ISR, the diffraction cone shows almost perfect exponential behaviour corresponding to a linear pomeron trajectory in a wide span of $0\lesssim|t|\lesssim1.3$ GeV$^2$, violated only by the "break" near $t= -0.1$~GeV$^2$, at the LHC it is almost immediately followed by another structure, namely by the dip at $t\approx -0.5$~GeV$^2$. The dynamic of the dip (diffraction minimum) has been treated fully and successfully \cite{Szanyi}, however those details are irrelevant to the behaviour of the pomeron trajectory in the resonance (positive $t$) region and expected glueballs there, that depend largely on the imaginary part of the trajectory and basically on the threshold singularity in Eq. (\ref{3}). 

In order to recover the pomeron and odderon trajectories in a widest possible interval of $t$, we fit the data including also the dip-bump and shoulder structures present in proton-proton and proton-antiproton scattering, respectively. 

At the TeV energy region, the pomeron and its $C$-odd counterpart, the odderon, entirely dominate \cite{Szanyi}, thus in this work we do not include the secondary reggeons into the fitting procedure.

A possible alternative to the simple Regge-pole model as input is a double pole (double pomeron pole, or simply dipole pomeron, DP) in the angular momentum $(j)$ plane \cite{PEPAN}:
\begin{eqnarray}\label{Pomeron}
& &A_P(s,t)={d\over{d\alpha_P}}\Bigl[{\rm e}^{-i\pi\alpha_P/2}G(\alpha_P)\Bigl(s/s_{0P}\Bigr)^{\alpha_P}\Bigr]= \\ \nonumber
& &{\rm e}^{-i\pi\alpha_P(t)/2}\Bigl(s/s_{0P}\Bigr)^{\alpha_P(t)}\Bigl[G'(\alpha_P)+\Bigl(L_P-i\pi
/2\Bigr)G(\alpha_P)\Bigr].
\end{eqnarray}
Since the first term in squared brackets determines the shape of the cone, one fixes
\begin{equation} \label{residue} G'(\alpha_P)=-a_P{\rm
	e}^{b_P[\alpha_P-1]},\end{equation} where $G(\alpha_P)$ is recovered
by integration. Consequently the pomeron amplitude Eq.~(\ref{Pomeron}) may be rewritten in the following "geometrical" form (for details see Ref.~\cite{PEPAN} and references therein):
\begin{equation}\label{GP}
A_P(s,t)=i{a_P\ s\over{b_P\ s_{0P}}}[r_{1P}^2(s){\rm e}^{r^{2}_{1P}(s)[\alpha_P-1]}-\varepsilon_P r_{2P}^2(s){\rm e}^{r^2_{2P}(s)[\alpha_P-1]}],
\end{equation} 
where $r_{1P}^2(s)=b_P+L_P-i\pi/2$, $r_{2P}^2(s)=L_P-i\pi/2$ and \mbox{$L_P\equiv\ln(s/s_{0P})$}. 

It has the advantages over the SP model that it produces logarithmically rising cross sections already at the "Born" level.


In earlier versions of the DP, to avoid conflict with the Froissart bound, the intercept of the pomeron was fixed at $\alpha_P(0)=1$. However later it was realized that the logarithmic rise of the total cross sections provided by the DP may not be sufficient to meet the data, therefore a supercritical intercept was allowed for. From the earlier fits to the data the value $\delta_P=\alpha_P(0)-1\approx0.04,$ half of Landshoff's value \cite{DDLN}, follows. This is understandable: the DP promotes half of the rising dynamics, thus moderating the departure from unitarity at the "Born" level (smaller unitarity corrections).

As in Ref.~\cite{Szanyi}, we assume that the odderon contribution is of the same form as that of the pomeron, implying the relation $A_O=-iA_P$ and different values of adjustable parameters (labeled by subscript ``$O$''): 
\begin{equation}\label{Odd}
A_O(s,t)={a_O\ s\over{b_O\ s_{0O}}}[r_{1O}^2(s){\rm e}^{r^2_{1O}(s)[\alpha_O-1]}-\varepsilon_O r_{2O}^2(s){\rm e}^{r^2_{2O}(s)[\alpha_O-1]}],
\end{equation}
where $r_{1O}^2(s)=b_O+L_O-i\pi/2$, $r_{2O}^2(s)=L_O-i\pi/2$, \mbox{$L_O\equiv\ln(s/s_{0O})$}. 

Both pomeron and odderon trajectories, $\alpha_P$ and $\alpha_O$, respectively, are defined by Eq.~(\ref{3}).

While the Pomeron has positive C-parity entering to the $pp$ and $\bar pp$ scattering amplitudes with the same sign, the Odderon has negative C-parity, entering to the $pp$ and $\bar pp$ scattering amplitudes with opposite sign:
\begin{equation}\label{Eq:Amplitude}
A\left(s,t\right)_{pp}^{\bar pp}=A_P\left(s,t\right)\pm A_O\left(s,t\right).
\end{equation}

We use the norm where
\begin{equation}\label{norm}
   \sigma_{tot}(s)={4\pi\over s}\Im m A(s,t=0)\Bigl.\ \ {\rm and}\ \
  {d\sigma_{el}\over{dt}}(s,t)={\pi\over s^2}|A(s,t)|^2 \  .
\end{equation}

The parameter $\rho(s)$, the ratio of the real to the imaginary part of the forward scattering amplitude is 
\begin{equation}\label{eq:rho}
\rho(s)=\frac{\Re e A(s,t=0)}{\Im m A(s,t=0)}.
\end{equation}

The elastic cross section $\sigma_{el}(s)$ is calculated by integration
\begin{equation}\label{eq:el}
\sigma_{el}(s)=\int_{t_{min}}^{t_{max}}\frac{d\sigma_{el}}{dt}(s,t)\, dt,
\end{equation}
whereupon
\begin{equation}\label{eq:inel}
\sigma_{in}(s)=\sigma_{tot}(s)-\sigma_{el}(s). 
\end{equation}
Formally, $t_{min}=-s/2$ and $t_{max}=t_{threshold}$, however since the integral is saturated basically by the first cone, we set $t_{max}=0$ and $t_{min}=-1$ GeV$^2$.

The free parameters of the model defined by the formulas Eqs.~(\ref{GP}-\ref{eq:rho}) were fitted simultaneously to the following dataset:

\begin{itemize}
	\item LHC TOTEM 7 and 13 TeV elastic $pp$ differential cross section data \cite{TOTEM13n,totem7} in the interval $0.005\leqslant|t|\leqslant3.8$ GeV$^2$;
	\item TEVATRON 1.8 and 1.96 TeV elastic $p\bar p$ differential cross section data \cite{Amos:1990fw,Abazov:2012qb} in the interval $0.03\leqslant|t|\leqslant1.2$ GeV$^2$;
	\item $pp$ and $p\bar p$ total cross section and $\rho$ parameter data \cite{totem13,TOTEM0,totem7,totem72,PDG,totem81,totem82} in the interval $10\leqslant\sqrt{s}\leqslant13000$ GeV.
\end{itemize}

Our main aim was to obtain the values of the parameters of the pomeron and odderon trajectories. The intercepts can be determined from the forward ($t=0$) data, while to determine the slope we need a fit in high-$|t|$ ranges. 

\begin{figure}[H] 
    \centering
    \includegraphics[scale=0.44]{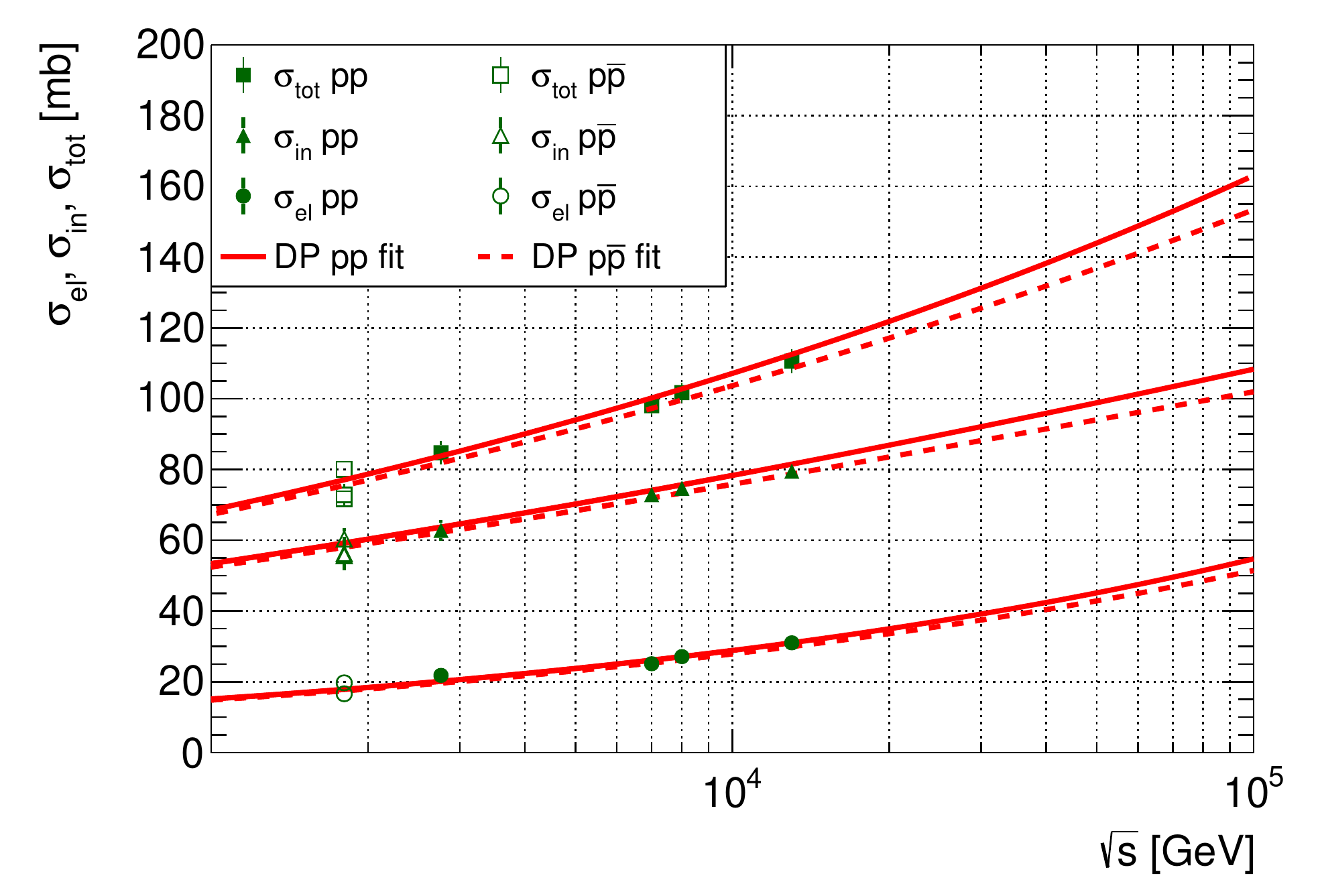}
    \caption{Fitted $pp$ and $p\bar p$ total cross section data with amplitude Eq.\ref{Eq:Amplitude} and trajectory Eq.~(\ref{3}) for pomeron and odderon. Calculated elastic and inelastic cross sections are also shown.}
    \label{fig:tot}
\end{figure}

\begin{figure}[H] 
	\centering
	\includegraphics[scale=0.44]{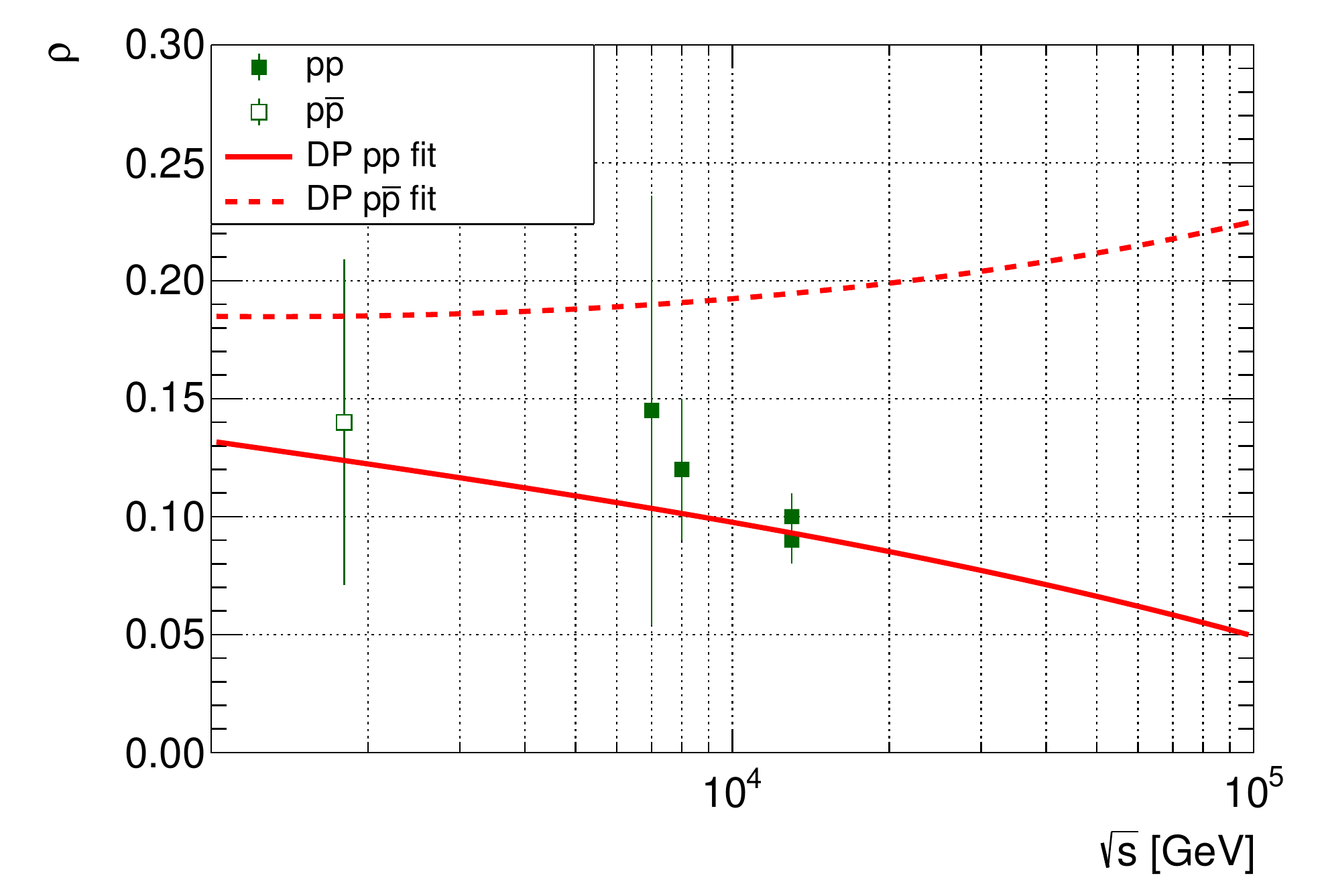}
	\caption{Fitted $pp$ and $p\bar p$ $\rho$-parameter data with amplitude Eq.\ref{Eq:Amplitude} and trajectory Eq.~(\ref{3}) for pomeron and odderon.}
	\label{fig:rho}
\end{figure}

\begin{figure}[H] 
	\centering
	\includegraphics[scale=0.44]{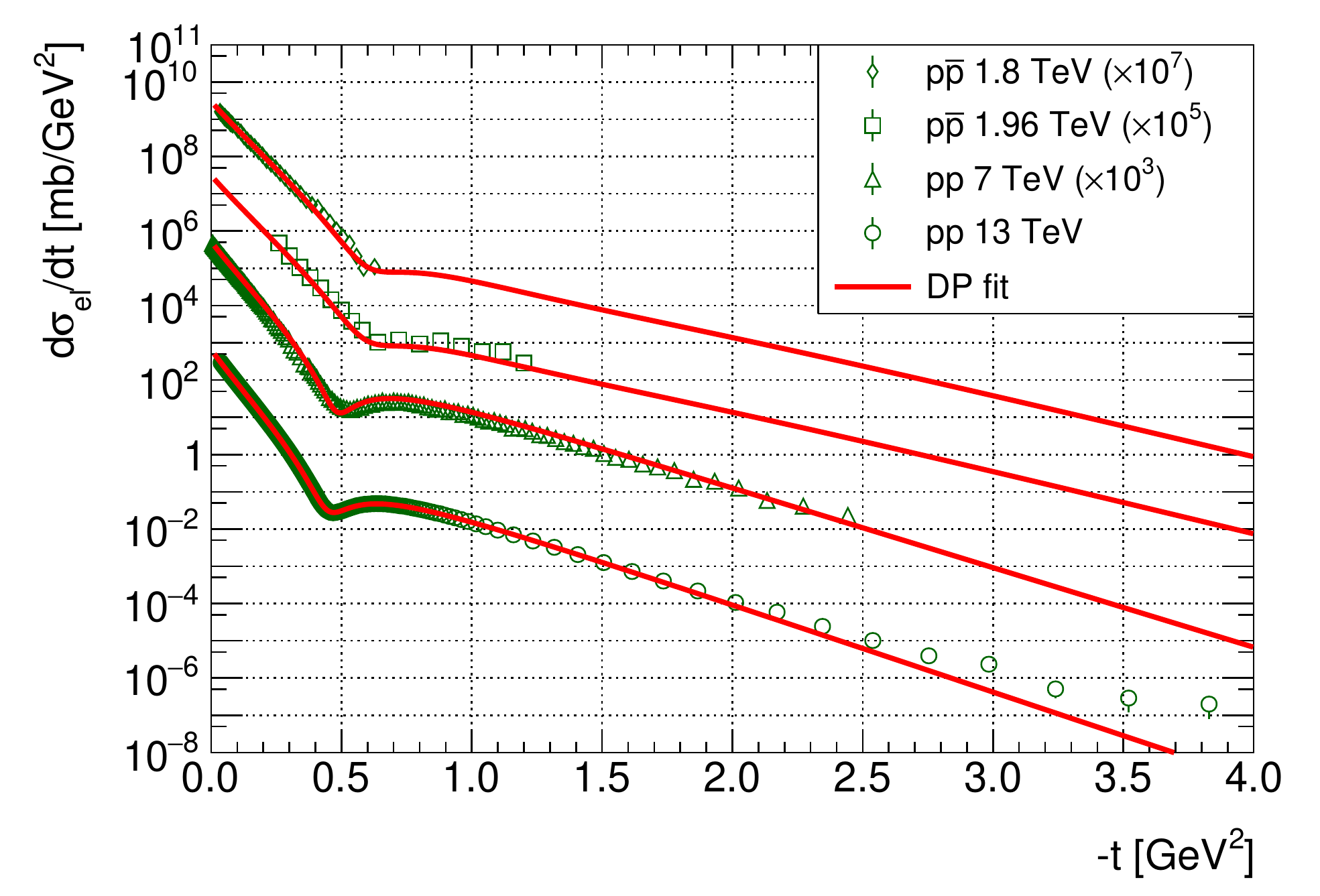}
	\caption{Fitted $pp$ and $p\bar p$ differential cross section data at 1.8, 1.96, 7 and 13 TeV with amplitude Eq.~(\ref{Eq:Amplitude}) and trajectory Eq.~(\ref{3}) for pomeron and odderon.}
	\label{fig:dsdt}
\end{figure}

The values of fitted parameters and the fit statistics are shown in Table~\ref{tab:parameters}. 
The results of the fits for the data on total cross section and $\rho$-parameter are shown in Figs.~\ref{fig:tot} and \ref{fig:rho}. The description for the differential cross section data is presented in Fig.~\ref{fig:dsdt}. 
The fit statistics indicates a need for improvement on the theoretical model side, however, as the figures show, the model catches all the features of the data except for the apparent slow-down of the exponential decrease of the cross section at the highest measured $|t|$ values which, as it was mentioned in Sec.~\ref{sec:unandu}, can be attributed to the transition from "soft" to "hard" dynamics. This behaviour could be mimicked by the gradual slow down (from linear to logarithmic) rise of the pomeron and/or odderon trajectories \cite{Jenk1}, however, this is beyond the scope of the present study as here we investigate only the "soft" dynamics. The value of the differential cross section where our fit begins to deteriorate is about the order of 10$^{-5}$. Thus this discrepancy between the fit and the measured data should not largely effect our predictions for the glueballs and oddballs.

Note that our fit, reasonably compatible with the measured data, results in a large odderon intercept, $\alpha_{O}(t=0)=1.23$. 
In our opinion, theoretical (QCD, lattice  etc.) calculations (see, e.g. Ref.\cite{LlanesEstrada:2005jf} and references therein) are rather formal, moreover, some predict an odderon intercept below $1$, while the odderon, like the pomeron, by definition is asymptotic, \textit{i.e.} its intercept should not fall \mbox{below 1.}

\begin{table}[H]    
	\centering
	\caption{Values of the parameters fitted to $pp$ and $p\bar p$ data on elastic differential cross section, total cross section and the ratio~$\rho$. Values of the fitted parameters and their errors are rounded up to four valuable decimal digits.}
	\begin{tabular}{|ccc|ccc|}
		\hline \hline
		\multicolumn{1}{|c}{}&\multicolumn{2}{c|}{Pomeron}&\multicolumn{2}{c}{Odderon}&\multicolumn{1}{c|}{} \\
		\hline 
		&$a_P~[\sqrt{\rm mbGeV^2}]$ & $318.9\pm2.123$ &$a_O~[\sqrt{\rm mbGeV^2}]$&$1.631\pm0.014$& \\
		&$b_P$ & $8.057\pm0.002$  &$b_O$ &$4.629\pm0.0178$&         \\
		&$\delta_P$ & $0.05180\pm0.00028$ &$\delta_O$&$0.2299\pm0.0004$&  \\
		&$\alpha_{1P}~[{\rm GeV}^{-2}]$ & $0.3051\pm0.0003$  &$\alpha_{1O}~[{\rm GeV}^{-2}]$ &$0.1951\pm0.0003$&  \\
		&$\alpha_{2P}~[{\rm GeV}^{-1}]$ & $0.07930\pm0.00045$  &$\alpha_{2O}~[{\rm GeV}^{-1}]$ &$0.03230\mp0.0002$&  \\
		&$\varepsilon_P$ & $0.4591\pm0.0016$  &$\varepsilon_O$&$2.982\pm0.004$&  \\
		&$s_{0P}~[{\rm GeV}^2]$ & $100$ (fixed) &$s_{0O}~[{\rm GeV}^2]$ & $100$ (fixed)&\\ \hline 
		\multicolumn{1}{|c}{}&\multicolumn{1}{c|}{Fit statistics}&\multicolumn{1}{c}{$\chi^2=1331.19$}&\multicolumn{1}{c}{$NDF=553$}&\multicolumn{1}{c}{$\chi^2/NDF=2.41$}&\multicolumn{1}{c|}{} \\ \hline \hline
	\end{tabular}
	\label{tab:parameters}
\end{table}

\section{Pomeron-pomeron and odderon-odderon total cross sections} \label{Amplitude}

In Ref. \cite{Schicker1, Schicker2} the resonances' contribution to pomeron-pomeron (PP) total cross section was calculated from the imaginary part of the amplitude by use of the optical theorem
\begin{align}\label{eq:imampl}
\sigma_{t}^{PP} (M^2) \;\; =& \;\; {\Im m\; A}(M^2, t=0) \;\;\\ \nonumber &  = \;\; a\sum_{i=f,P}\sum_{J}\frac{[f_{i}(0)]^{J+2}\; \Im m \;\alpha_{i}(M^{2})}
{\left[J-\Re e \;\alpha_{i}(M^{2})\right]^{2}+\left[\Im m \;\alpha_{i}(M^{2})\right]^{2}}.
\end{align}

Considering only the pomeron component, Eq.~(\ref{eq:imampl}) reduces to
\begin{equation}
\sigma_{t}^{PP} (M^2) \; = \; a\sum_{J}\frac{k^{J+2}\; \Im m \;\alpha_P(M^{2})}
{\left[J-\Re e \;\alpha_P(M^{2})\right]^{2}+\left[\Im m \;\alpha_P(M^{2})\right]^{2}},
\label{eq:imamplP}
\end{equation}
where $k=f_P(0)$, and, for simplicity here we set $k=1$. The numerical value of the prefactor $a$ in Eq.~(\ref{eq:imamplP}) is 1 GeV$^{-2}$ = 0.389 mb.

We use the same formula Eq.~(\ref{eq:imamplP}) for the odderon component of the odderon-odderon (OO) total cross section changing the indexes from "$P$" to "$O$".

\begin{figure}[H]
	\centering
	\includegraphics[scale=0.44]{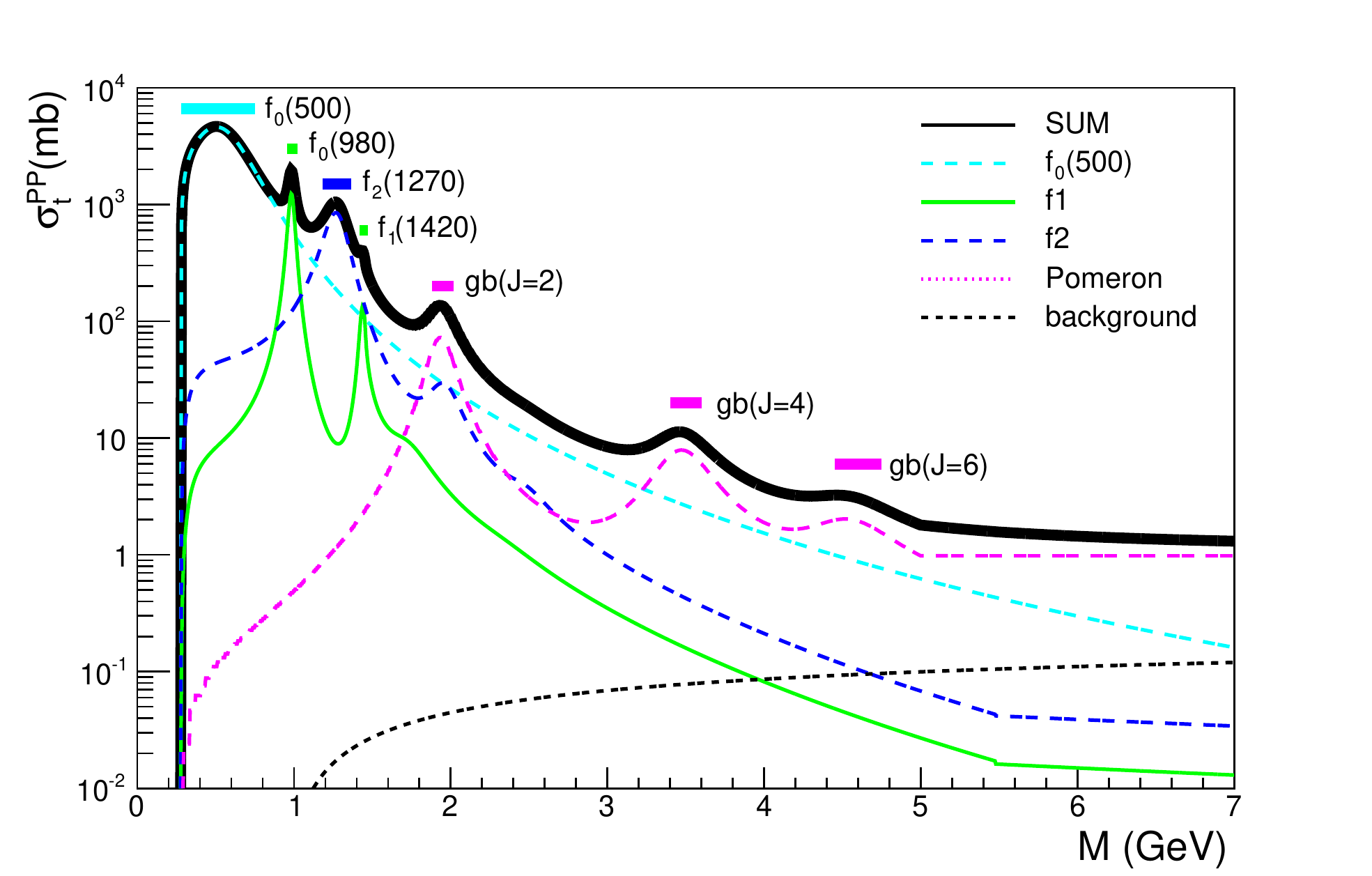}
	\caption{Pomeron-pomeron total cross section in CED, calculated in Ref.~\cite{trento}.}
	\label{fig:FJSch}
\end{figure}

Preliminary results on pomeron-pomeron total cross section,  Fig.\ref{fig:FJSch}, using the new trajectory, Eq.~(\ref{3}), were presented in Ref.~\cite{trento} and more recently in Ref.~\cite{Szanyi:2019pye}.

In this work we start by investigating the resulting glueball/oddball spectra in two ways: first we plot the real and imaginary parts of the trajectories (Chew-Frautschi plot) and calculate the resonances' widths by using the relation (see, \textit{e.g.} Eq. (18) in Ref.~\cite{Francesco_b})
\begin{equation}\label{eq:Chew_F}
\Gamma(t=M^2)=\frac{2\Im m \alpha(M^2)}{|\alpha_R'(M)|}
\end{equation}
where $\alpha'(M)=dRe\alpha(M)/dM$. Then we calculate the pomeron/odderon component of the PP/OO total cross section by using the Breit-Wigner formula, Eq. (\ref{eq:imamplP}). 

\section{Glueballs}\label{sec:glueballs}
\begin{figure}[H] 
	\centering
	\includegraphics[scale=0.44]{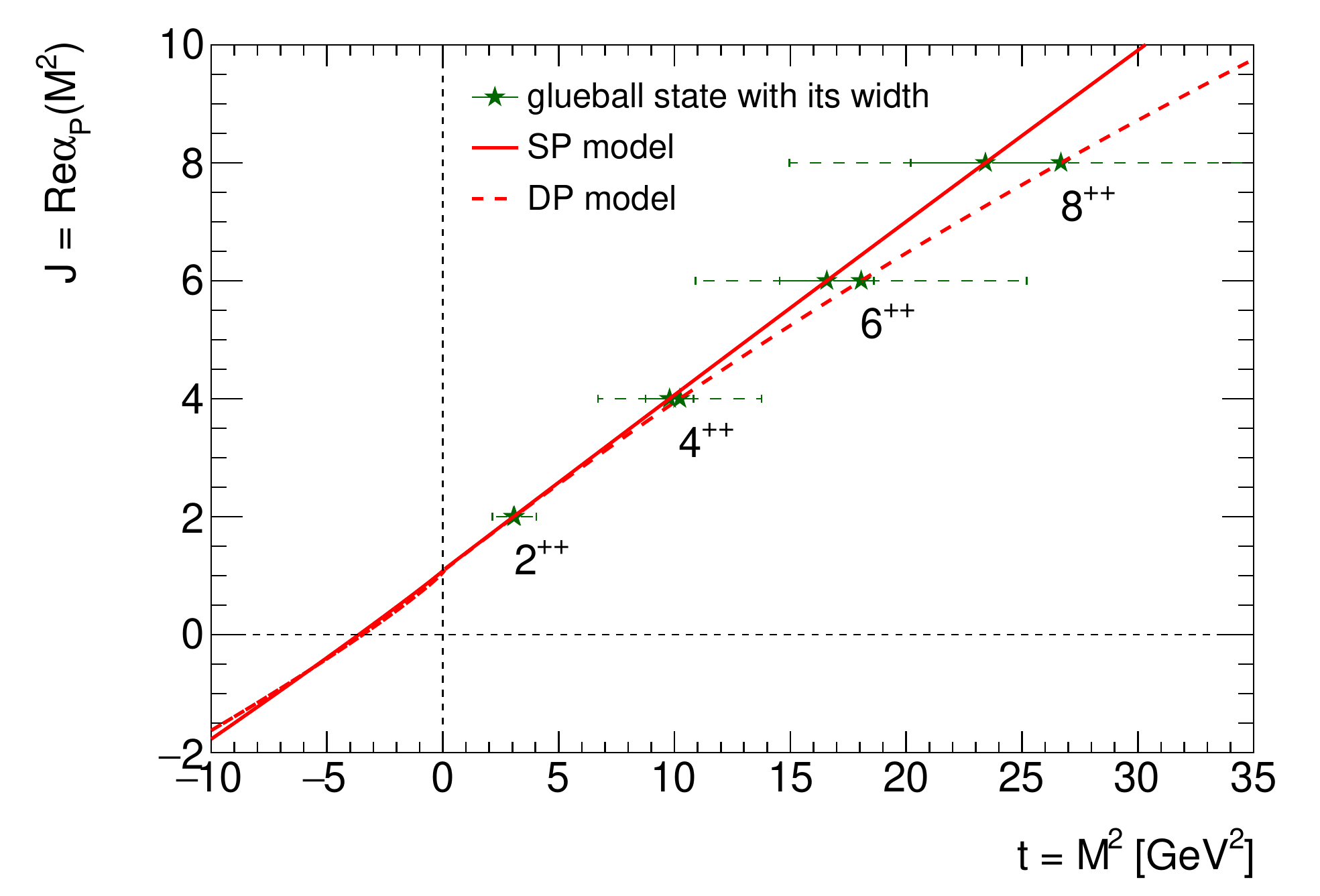}
	\caption{Real part of the pomeron trajectory Eq.~(\ref{3}) both for SP and DP model cases. The widths of resonances (glueball states) are shown in form of horizontal bars.}
	\label{fig:re}
\end{figure}

\begin{figure}[H] 
	\centering
	\includegraphics[scale=0.44]{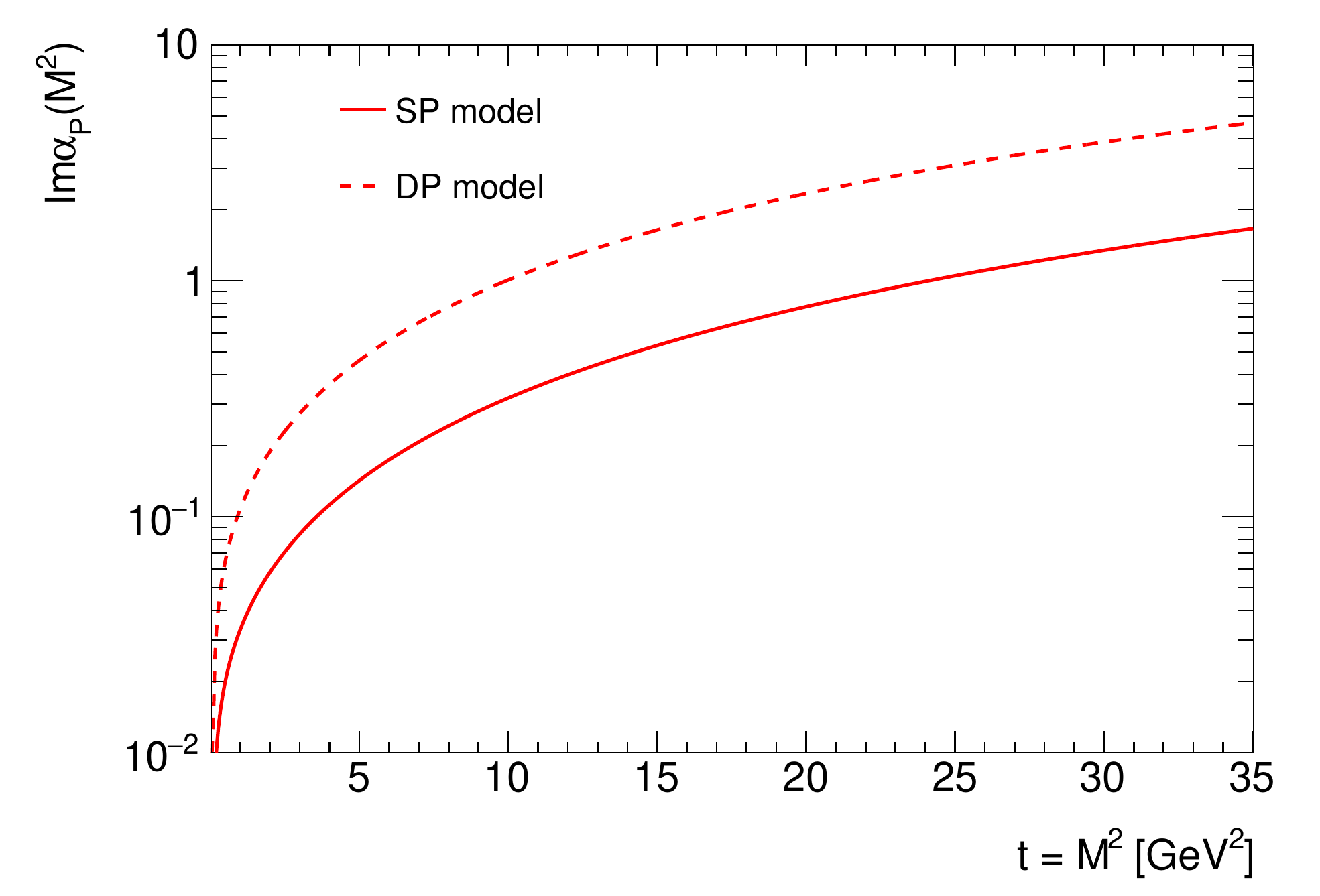}
	\caption{Imaginary parts of the pomeron trajectory Eq.~(\ref{3}) both for SP and DP model cases.}
	\label{fig:im}
\end{figure}

\begin{figure}[H] 
	\centering
	\includegraphics[scale=0.44]{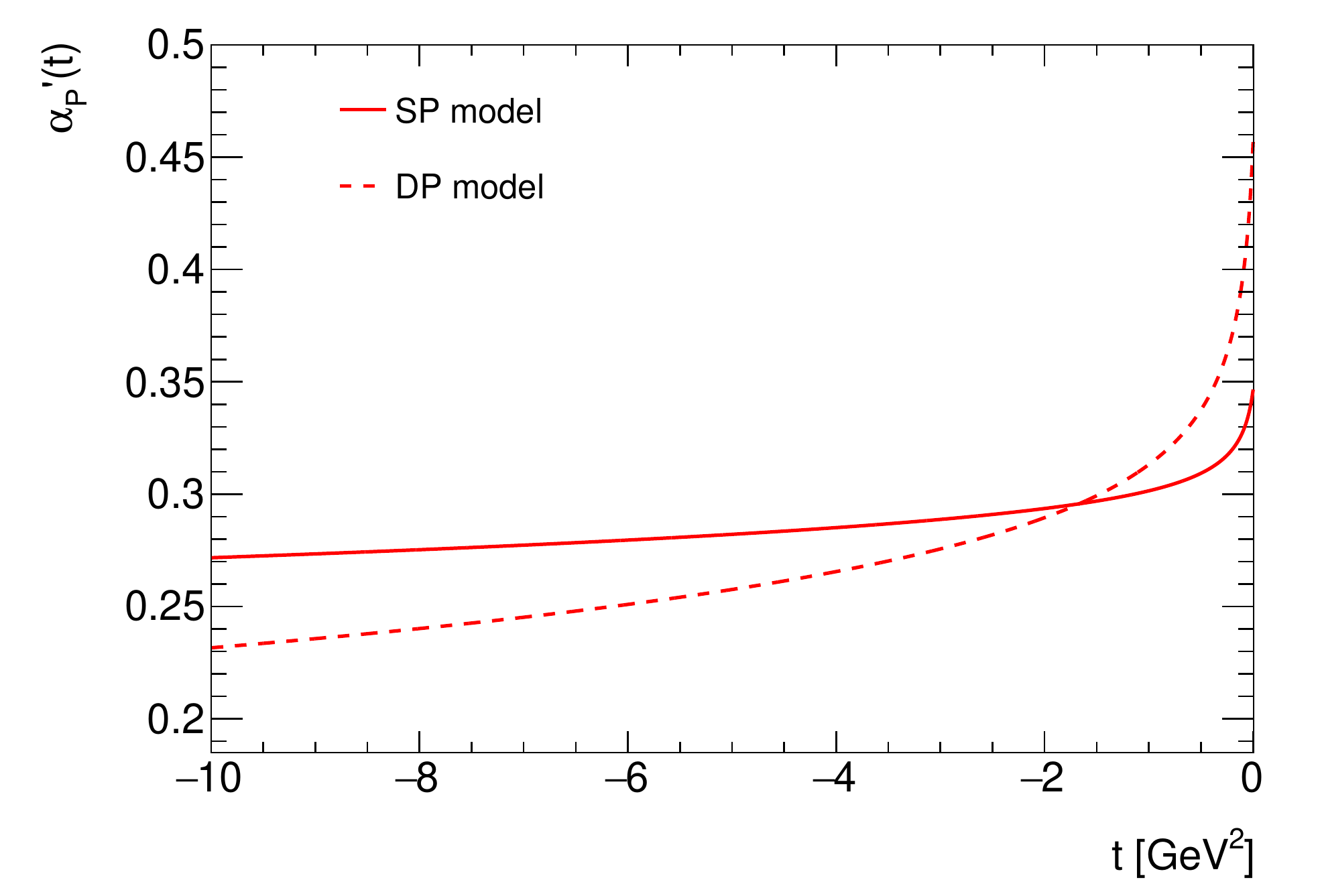}
	\caption{Slope of the pomeron trajectory Eq.~(\ref{3}) both for SP and DP model cases.}
	\label{fig:slope}
\end{figure}

With sets of parameters, taken from the fits of both SP and DP models, we extrapolate the pomeron/glueball trajectory to the resonance region. In the resonance region by realizing crossing symmetry $t$ becomes $s$ giving the mass squared $M^2$ of the resonances. The real and imaginary parts of the pomeron trajectory obtained by the SP and DP fits are shown in Figs.~\ref{fig:re} and \ref{fig:im}. Fig.~\ref{fig:re} shows also the predicted glueball states (with their widths) lying on the pomeron trajectory. These states are summarized in Tabs.~\ref{tab:pred_gluebs_SP} and \ref{tab:pred_gluebs_DP}. The $t$-dependence of the slope of the pomeron trajectory is shown in Fig.~\ref{fig:slope}. 

One can see that the widths of resonances are not the same for SP and DP fits. The advantage of the DP model that it is applicable in a broader $t$ range and gives the possibility to predict also the oddball states. However, the DP produces a faster growth for the slope of the trajectory than that of the SP model, which originally should be almost constant.

\begin{figure}[H] 
	\centering
	\includegraphics[scale=0.44] {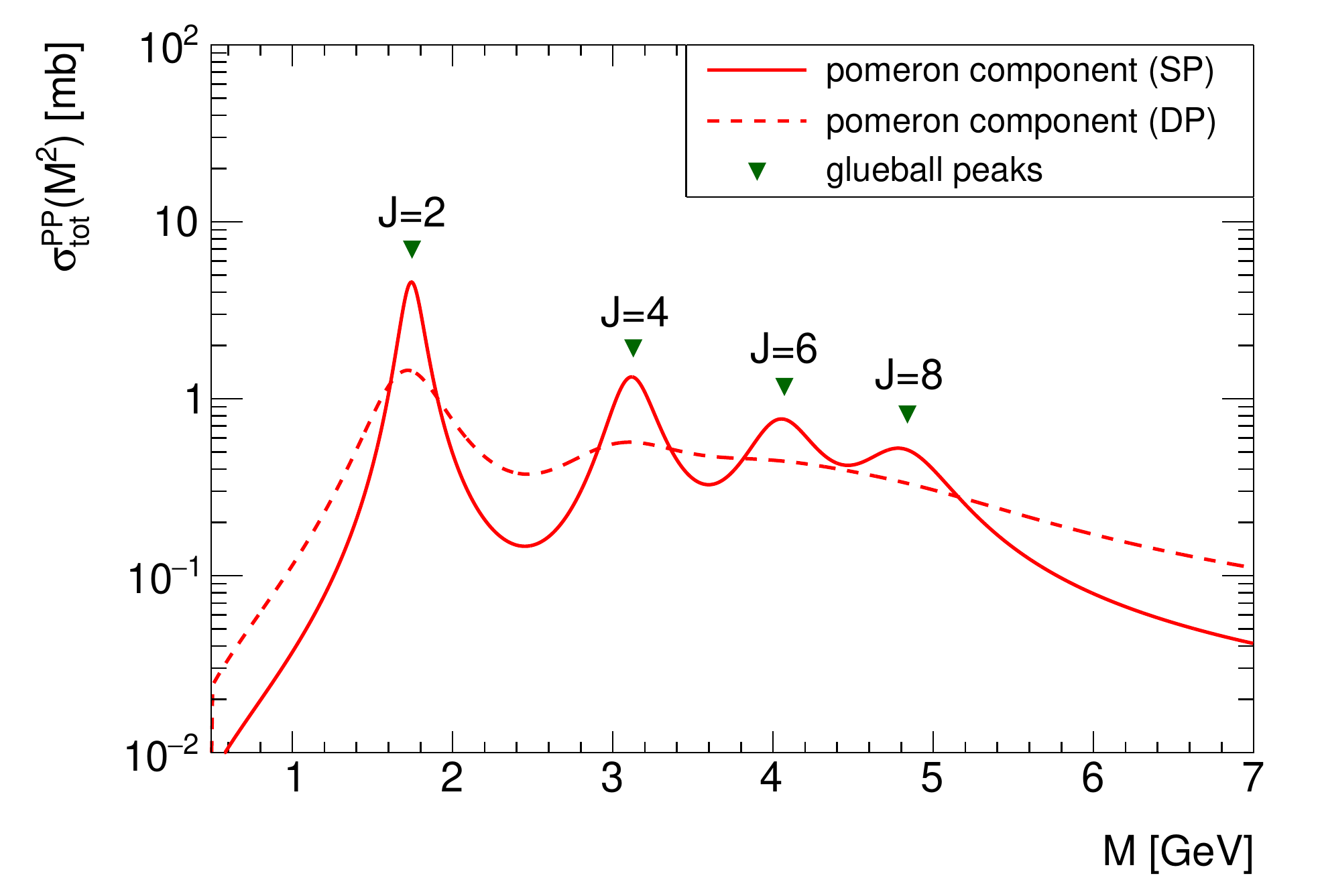}	\caption{The pomeron component of PP total cross section Eq.~(\ref{eq:imamplP}) with trajectory Eq.~(\ref{3}) both for SP and DP model cases. }
	\label{fig:sigma_ratio}
\end{figure}

The predicted pomeron component of PP total cross section both for SP and DP model cases are shown in Fig.~\ref{fig:sigma_ratio}. 

\begin{table}[H]    
	\centering
	\caption{Masses and widths of the glueball states predicted by using the SP Regge model.}
	\begin{tabular}{|c|c|c|}
		\hline
		$J^{PC}$& $M$, GeV & $\Gamma$, GeV  \\ \hline \hline
		$2^{++}$& $1.747$ & $0.574$  \\
		$4^{++}$& $3.128$ & $2.076$ \\
		$6^{++}$& $4.071$ & $4.064$\\
		$8^{++}$& $4.839$ & $6.452$\\
		\hline
	\end{tabular}
	\label{tab:pred_gluebs_SP}
\end{table}

\begin{table}[H]    
	\centering
	\caption{Masses and widths of the glueball states predicted by using the DP Regge model.}
	\begin{tabular}{|c|c|c|}
		\hline
		$J^{PC}$& $M$, GeV & $\Gamma$, GeV  \\ \hline \hline
		$2^{++}$& $1.758$ & $1.888$  \\
		$4^{++}$& $3.198$ & $7.063$ \\
		$6^{++}$& $4.249$ & $14.292$\\
		$8^{++}$& $5.165$ & $23.452$\\
		\hline
	\end{tabular}
	\label{tab:pred_gluebs_DP}
\end{table}

\section{Oddballs}\label{sec:oddballs}

Oddballs, resonances made of three gluons, have the same right of existence as glueballs made of two gluons. Oddballs are expected to lie on the odderon trajectory exactly in the same way as glueballs lie on the pomeron trajectory. The problem is that the odderon contribution to proton-proton and proton-antiproton scattering in the nearly forward direction is suppressed at least by an order of magnitude with respect to that of the pomeron. 

Anyway, one may try, given the existing parametrization of the odderon. They are available \textit{e.g.} in \cite{Tan,Szanyi}, where, however, for simplicity a linear trajectory (uninteresting for the present purposes) was used. Above we presented an update of the fits of Refs.~\cite{Tan, Szanyi} with the threshold singularity included in the odderon trajectory. Note that in the case of odderon exchange the lowest threshold is at $t_{0O}=(3m_{\pi})^2$, and not at $(2m_{\pi})^2$ as in case of the pomeron.

In predicting masses and decay widths of oddballs lying on the odderon trajectory, we use the result of our fit presented in Sec.~\ref{Sec:DP}, in Tab.~\ref{tab:parameters}.

\begin{figure}[H] 
	\centering
	\includegraphics[scale=0.44]{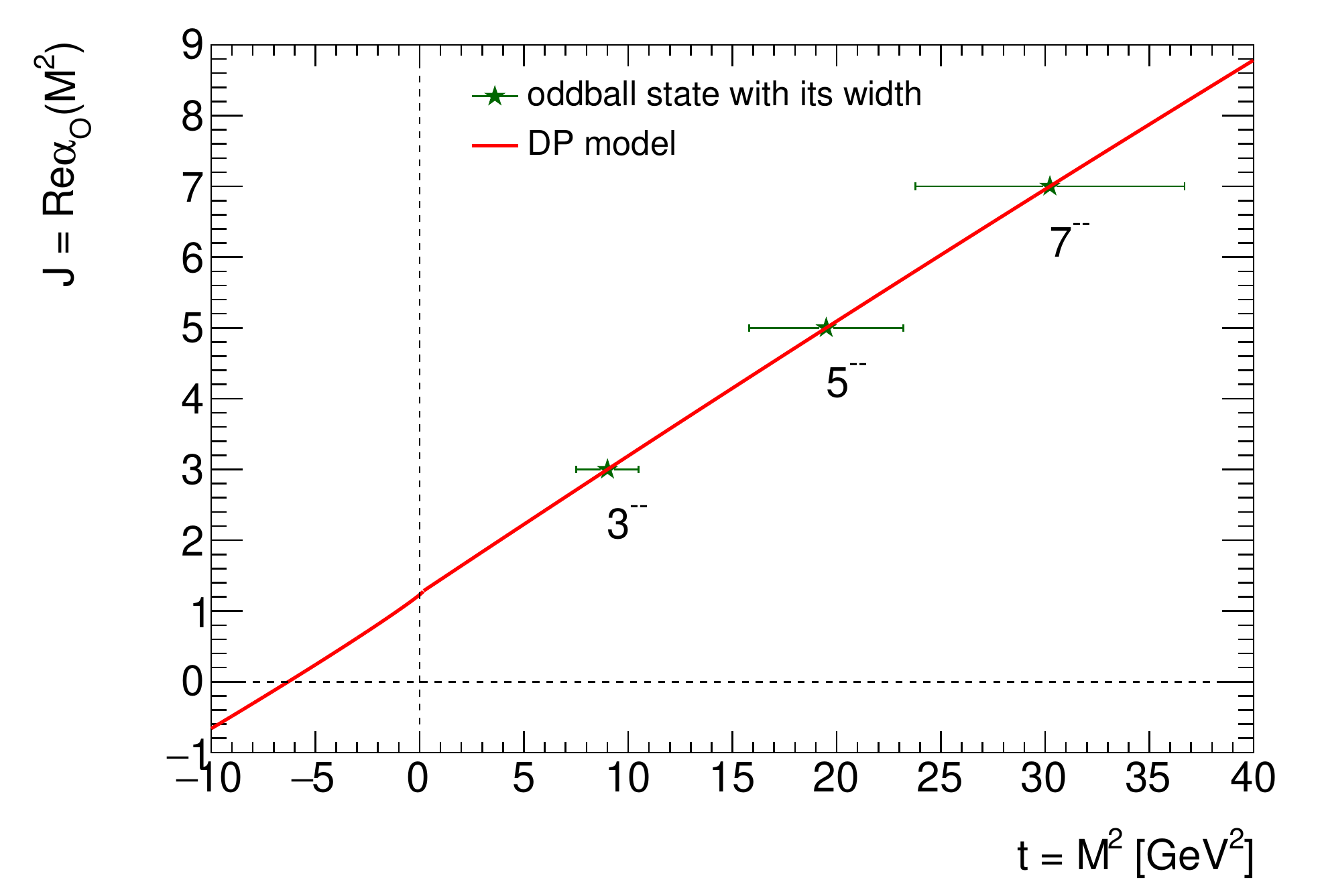}
	\caption{Real part of the odderon trajectory Eq.~(\ref{3}). The widths of resonances (oddball states) are shown in the form of horizontal bars.}
	\label{fig:re_odderon}
\end{figure}

\begin{figure}[H] 
	\centering
	\includegraphics[scale=0.44]{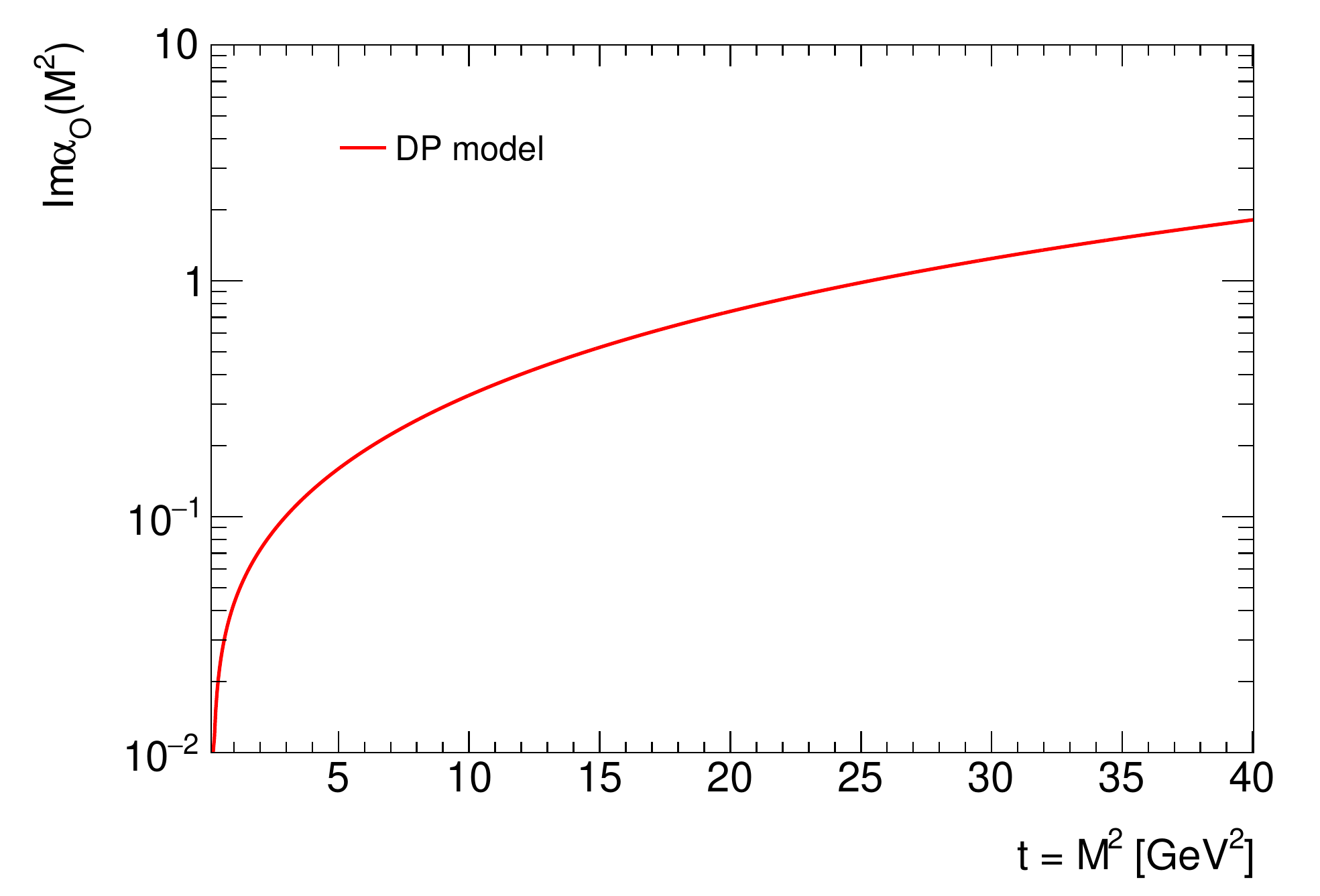}
	\caption{Imaginary part of the odderon trajectory Eq.~(\ref{3}).}
	\label{fig:im_odderon}
\end{figure}

\begin{figure}[H] 
	\centering
	\includegraphics[scale=0.44]{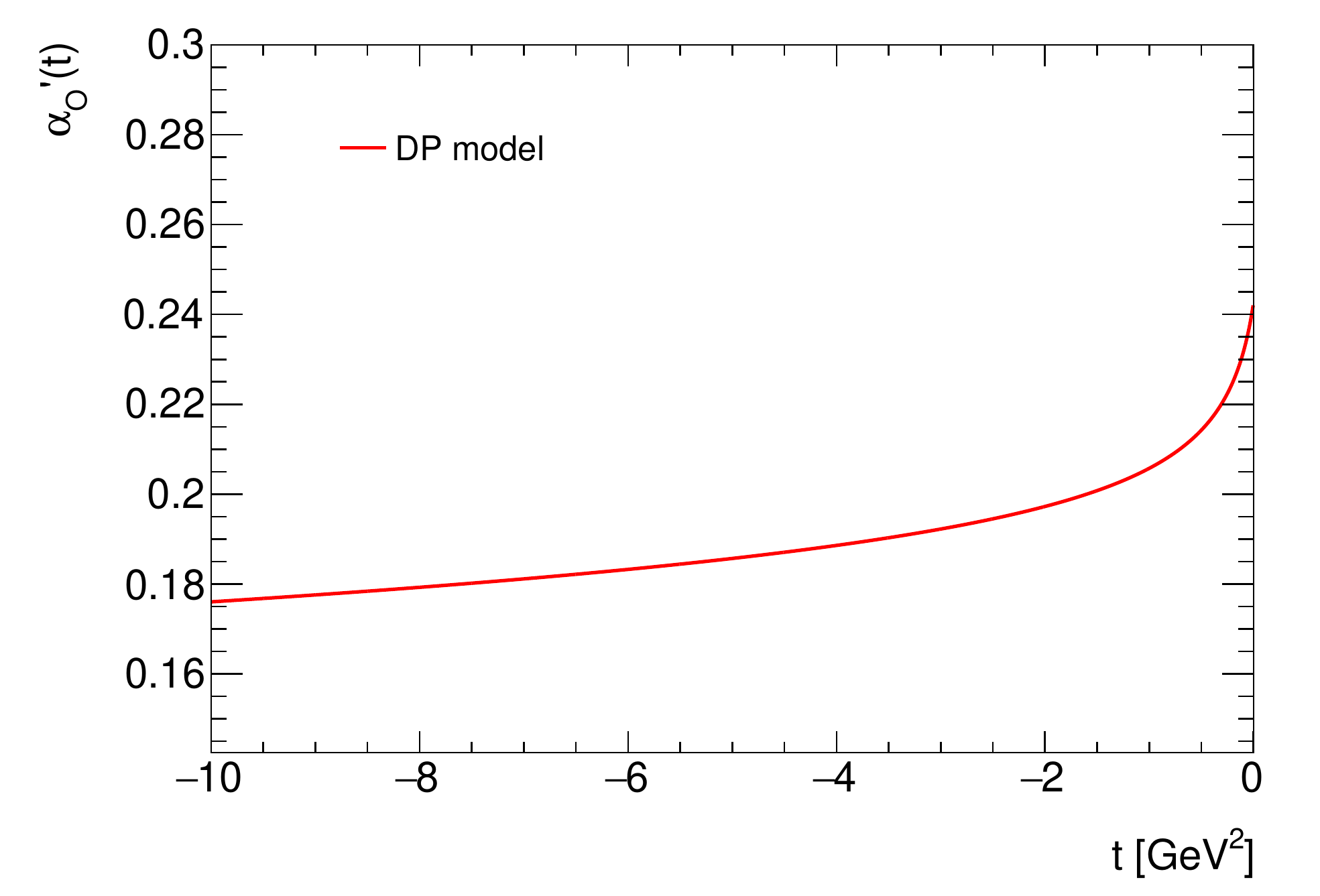}
	\caption{Slope of the odderon trajectory Eq.~(\ref{3}).}
	\label{fig:slope_odderon}
\end{figure}

\begin{figure}[H] 
	\centering
	\includegraphics[scale=0.44]{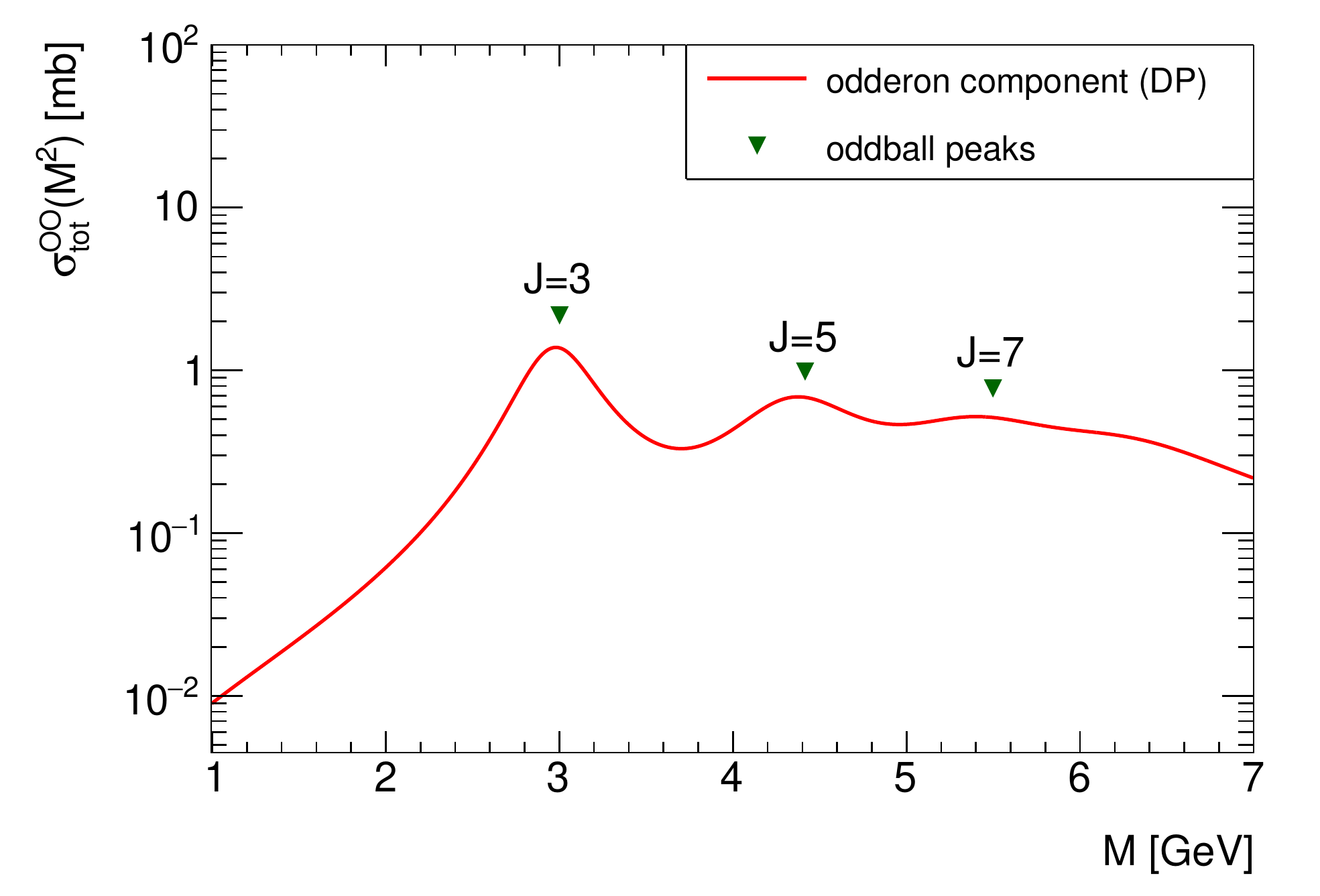}
	\caption{The odderon component of OO total cross section calculated with trajectory Eq.~(\ref{3}).}
	\label{fig:width_cross_odderon}
\end{figure}

The real and imaginary parts of the odderon trajectory are shown in Figs.~\ref{fig:re_odderon} and \ref{fig:im_odderon}. Fig.~\ref{fig:re_odderon} shows also the predicted oddball states (with their widths) lying on the odderon trajectory. These states are summarized in Tab.~\ref{tab:pred_oddbs_DP}. The energy dependence of the slope of the odderon trajectory is shown in Fig.~\ref{fig:slope_odderon}. The predicted odderon component of OO total cross section is shown in Fig.~\ref{fig:width_cross_odderon}.

\begin{table}[H]    
	\centering
	\caption{Masses and widths of the oddball states predicted by using the DP Regge model.}
	\begin{tabular}{|c|c|c|}
		\hline
		$J^{PC}$& $M$, GeV & $\Gamma$, GeV  \\ \hline \hline
		$3^{--}$& $3.001$ & $2.984$  \\
		$5^{--}$& $4.416$ & $7.379$ \\
		$7^{--}$& $5.498$ & $12.913$\\
		\hline
	\end{tabular}
	\label{tab:pred_oddbs_DP}
\end{table}
    
\section{Conclusions}\label{sec:conclusions}

In this paper we rely on the $S$-matrix theory, in particular on the Regge-pole model, free of any problem around $t=0$. At the same time, being aware of the discrepancy with QCD and the string theory at this point, we do not exclude that the non-observability of glue- and oddballs are connected with the loss of analyticity in the QCD and string theory at the “special point”, $t=0$, an interesting point to be investigated.
 
We have presented a new, simple model for Regge trajectories producing physical resonances with increasing widths but satisfying also threshold behaviour required by unitarity and asymptotic behaviour compatible with the bounds imposed by analyticity and duality. The novelty of this model that it produces resonances with increasing widths. This trajectory models the pomeron and odderon. It has only three free parameters. One of them ($\delta$) is uniquely related to the intercept while the other two ($\alpha_1$ and $\alpha_2$) mainly define the slope of the trajectory. These parameters were fitted to the high-energy proton-proton and proton-antiproton scattering data. We have performed two fits: one to the LHC $pp$ data at low-$|t|$, below the dip, appearing near $|t|\approx 0.5$ GeV$^2$ and another one, including all available data, \textit{i.e.} up to $|t|
\approx 4$ GeV$^2.$
The first fit has the advantage to follow the simple Donnachie-Landshoff model (a single power), however it is strongly affected by the "break" near $|t|\approx 0.1$ GeV$^2$.  

The subsequent (beyond the break) nearly linear behaviour of the trajectory (and exponential cone), was restored in our second fit. The cone is interrupted by the dip, but structureless beyond the dip.  Note that while the "break" comes mainly from the trajectory, the dip-bump structure is a property of the amplitude rather than of the trajectory. By using a successful model for the amplitude producing the dip and bump, independent of the form of the trajectory, we have a chance to restore the trajectories (of the pomeron and odderon) in a wide span of $t$. 

Interestingly, as it was shown in Ref.~\cite{Szanyi}, in the second cone (beyond the dip-bump) the odderon takes over the pomeron, compatible with Landshoff's idea \cite{Landshoff} about three gluon exchange at the second cone.

Towards largest available values of $|t|$ the exponential decrease of the cross section tends to slow down toward a power behaviour corresponding to the onset of the large-$|t|$ logarithmic asymptotics of the trajectories producing wide-angle scaling behaviour of the cross sections \cite{Rivista,Jenk1}, that goes beyond the scope of the present paper. 

More relevant for future studies is the possibility to apply the new trajectory Eq.~(\ref{3}) to ordinary reggeons and known mesons and baryons. 

Given the nearly linear behaviour of the pomeron (and odderon) trajectory (constrained by the observed diffractive peak), at low and moderate values of its parameter, the predicted masses of glueballs (and oddballs) are consistent with similar predictions based on linear trajectories. Non-trivial are predictions of glueballs' (and oddballs') widths. Our Fig.~\ref{fig:sigma_ratio} shows that even small variations of the parameters in the trajectory result in noticeable changes of glueballs' widths.

Finally, let us stress that the new trajectory is an important building block in the CED program to be continued along the lines of paper \cite{Schicker1, Schicker2, Ciesielski}, see the Appendix.  

\subsection*{Acknowledgements}
L.~J. thanks A. Bugrij for useful discussions. L.~J. was supported by the Ukrainian Academy of Sciences' program "Structure and dynamics of statistic and dynamic quantum-mechanical systems". The work of I.~Sz. was supported by the "M\'arton \'Aron Szakkoll\'egium" program. The work of R.S. is supported by the German Federal Ministry of Education and  Research under promotional reference 05P19VHCA1. We thank Felipe J. Llanes Estrada for a useful correspondence. 

We thank the Referee for the remarks and recommendations, which definitely helped us to improve the paper.

\section*{Appendix}  

For sufficiently large rapidity gaps the cross sections may be written as \cite{ Ciesielski,Dino},  
\begin{eqnarray}
\frac{d^2\sigma_{SD}}{dtd\Delta y}\!\!&\!\!=\!\!&\!\! \frac{1}{N_{\rm gap}(s)}\!\!\left[ \frac{~ ~ \beta^2(t)}{16\pi}e^{2[\alpha(t)-1]\Delta y}\right] \cdot \left\{ \kappa \beta^2(0) \left( \frac{s'}{s_{0}}\right)^{\epsilon}\right\}, \label{eqSD}\\
\frac{d^3\sigma_{DD}}{dtd\Delta y dy_0}\!\!&\!\!=\!\!&\!\! \frac{1}{N_{\rm gap}(s)}\!\!\left[ \frac{\kappa\beta^2(0)}{16\pi}e^{2[\alpha(t)-1]\Delta y}\right] \cdot \left\{ \kappa \beta^2(0) \left( \frac{s'}{s_{0}}\right)^{\epsilon}\right\}, \label{eqDD}\\
\frac{d^4\sigma_{DPE}}{dt_1dt_2d\Delta y dy_c}\!\!&\!\!=\!\!&\!\! \frac{1}{N_{\rm gap}(s)}\!\!\left[\Pi_i\!\!\left[ \frac{\beta^2(t_i)}{16\pi}e^{2[\alpha(t_i)-1]\Delta y_i}\right]\!\right]\!\!\cdot\!\kappa\!\left\{ \kappa \beta^2(0)\!\!\left( \frac{s'}{s_{0}}\right)^{\!\!\epsilon}\right\}, \label{eqCD}
\end{eqnarray}
where $t$ is the square of the four-momentum-transfer at the proton vertex and $\Delta y$ is the rapidity gap width. The variable $y_0$ in Eq.~(\ref{eqDD}) is the center of the rapidity gap. In Eq.~(\ref{eqCD}), the subscript $i=1, 2$ enumerates pomerons in the DPE event, $\Delta y=\Delta y_1 + \Delta y_2$ is the total (sum of two gaps) rapidity-gap width in the event, and $y_c$ is the center in $\eta$ of the centrally-produced hadronic system.

Eqs.~(\ref{eqSD}) and (\ref{eqDD}) are equivalent to those of standard Regge theory, as $\xi$, the fractional forward-momentum-loss of the surviving proton (forward momentum carried by pomerons), is related to the rapidity gap by $\xi=e^{-\Delta y}$. 
The variable $\xi$ is defined as $\xi_{\rm SD}=M^2/s$ and $\xi_{\rm DD}=M_1^2M_2^2/(s\cdot s_0)$, where $M^2$ ($M_1^2$, $M_2^2$) are the masses of dissociated systems in SD (DD) events. For DD events, $y_0=\frac{1}{2}\ln(M_2^2/M_1^2)$, and for  DPE $\xi=\xi_1\xi_2=M^2/s$.

The pomeron-proton ($P$-$p$) coupling , $\beta(t)$, is given by:
\begin{equation}
\beta^2(t)=\beta^2(0)F^2(t),
\end{equation}
where $\beta(0)= 4.0728 ~ \sqrt{\mbox{mb}} = 6.566 ~ \mbox{GeV}^{-1}$ and $F(t)$ is the residue function, that in Ref. \cite{Ciesielski} was identified with the proton form factor:
\begin{equation}
F^2(t)=\left[ \frac{4m^2_p-2.8t}{4m^2_p-t}\left(\frac{1}{1-\frac{t}{0.71}} \right)^2\right]^2 \approx a_1e^{b_1t}+a_2e^{b_2t}.
\label{eqFF}
\end{equation}
An alternative was advocated in \cite{Tan}.
The right-hand side of Eq.~(\ref{eqFF}) is a double-exponential approximation of $F^2(t)$, with $a_1=$0.9, $a_2$=0.1, $b_1=$4.6 GeV$^{-2}$, and $b_2=$0.6 GeV$^{-2}$.
 The term in curly brackets in Eqs.(\ref{eqSD})-(\ref{eqCD}) 
is the $P$-$p$ total cross section at the reduced $P$-$p$ collision energy squared, $s'=s \cdot e^{-\Delta y}$. 
The parameter $\kappa$ is set to $\kappa=0.17$~\cite{Ciesielski},  and $\kappa \beta^2(0)\equiv \sigma_0$, where $\sigma_o$ defines the total pomeron-proton cross section at an energy-squared value  of $s_0 = 1~\mbox{GeV}^2$, is set to $\sigma_0= 2.82~\mbox{mb}$ or $7.249~\mbox{GeV}^{-2}$.
 
The expressions in curly brackets above are Regge asymptotic total cross sections, reflecting the smooth behaviour of pomeron-pomeron scattering at high missing masses \cite{Schicker2}.

\section*{References}
\bibliography{mybibfile}
\end{document}